\documentstyle{amsart}

\numberwithin{equation}{section}

\newtheorem{theorem}{Theorem}[section]

\newtheorem{lemma}[theorem]{Lemma}
\newtheorem{corollary}[theorem]{Corollary}
\newtheorem{proposition}[theorem]{Proposition}

\theoremstyle{remark}
\newtheorem{remark}[theorem]{Remark}   
\newtheorem{remarks}{Remarks}

\newcommand{\diag}{\operatorname{diag}}
\newcommand{\Ad}{\operatorname{Ad}}
\newcommand{\tr}{\operatorname{tr}}
\newcommand{\Res}{\operatorname{Res}}
\newcommand{\real}{\operatorname{Re}}
\newcommand{\imag}{\operatorname{Im}}

\begin{document}

\title[Asymptotic metrics for $SU(N)$-monopoles]
      {Asymptotic metrics for $SU(N)$-monopoles with maximal symmetry breaking}
\author{Roger Bielawski}

\address{ Max-Planck-Institut f\"ur Mathematik\\ Gottfried-Claren-Strasse 26\\
53225 Bonn, Germany}

\email{rogerb@@mpim-bonn.mpg.de}

\begin{abstract} We compute the asymptotic metrics for moduli spaces of $SU(N)$ monopoles with maximal symmetry breaking. These metrics are exponentially close to the exact monopole metric $\!^1$ as soon as, for each simple root, the individual monopoles corresponding to that root are well separated. We also show that the estimates can be differentiated term by term in natural coordinates, which is a new result even for $SU(2)$ monopoles. 
      
\end{abstract}

\maketitle

\section{Introduction\label{-one}}

It has been known since the work of Taubes \cite{Taubes} that an $SU(2)$-monopole of charge $m$ with well-separated zeros of the Higgs field approximates a collection of $m$ monopoles of charge $1$. This fact is reflected in the asymptotic behaviour of the natural hyperk\"ahler metric on the moduli space $M_m$ of charge $m$ monopoles. Namely, in the asymptotic region of $M_m$, the monopole metric is exponentially close to another hyperk\"ahler metric, whose geodesics determine scattering of $m$ particles with electric, magnetic and scalar charges. This metric was found by Gibbons and Manton in \cite{GM} and a proof that it differs from the exact monopole metric by an exponentially small amount was given in \cite{BielCMP}. For $m=2$, the Gibbons-Manton metric is just the product of a flat metric and the Taub-NUT metric with a negative mass parameter.
\par
Monopoles exist for any compact Lie group and Taubes' estimates work equally well for $SU(N)$-monopoles with maximal symmetry breaking, that is monopoles whose Higgs field has distinct eigenvalues at infinity. This time a moduli space $M_{m_1,\dots,m_{N-1}}(\mu_1,\dots,\mu_N)$, where $m_i$ are positive integers and $\mu_1<\dots<\mu_N$, is obtained by identifying gauge-equivalent framed monopoles whose Higgs field at infinity defines a map from the $2$-sphere to the adjoint orbit $O$ of $\diag(i\mu_1,\dots,i\mu_N)$ and whose degree is $(m_1,\dots,m_{N-1})\in H_2(O,{\Bbb Z})$. We should think of particles making up the monopole as coming in $N-1$ distinguishable types, with $m_i$ being the number of particles of type $i$.
\par
In this paper we shall compute the asymptotic metric on $M_{m_1,\dots,m_{N-1}}(\mu_1,\dots,\mu_N)$ \nolinebreak \footnote{Strictly speaking, we only show that our asymptotic metric is close to the metric on the moduli space of solutions to Nahm's equations. Thus we can conclude that the geodesics of the monopole metric and of the asymptotic metric are close to each other (see Remark \ref{isometry}). The metrics themselves are close if the Nahm transform for $SU(N)$-monopoles, $N>2$, is an isometry.}. \linebreak This metric turns out to be a hybrid between the Gibbons-Manton metric and the metric on the moduli space of monopoles of charge $(1,1,\dots,1)$ which was computed in \cite{LWY,MM,Cha}. Particles of the same type interact as in the Gibbons-Manton metric while the particles of different types as in the $(1,1,\dots,1)$-metric (i.e. neighboring types interact as in the Taub-NUT metric with a positive mass parameter and non-neighboring types do not interact). The precise formula for the asymptotic metric is given in the next section. Somewhat surprisingly, the monopole metric on $M_{m_1,\dots,m_{N-1}}(\mu_1,\dots,\mu_N)$ is exponentially close to the asymptotic metric as soon as, for each $i$, particles of type $i$ are far apart. Particles of different types can be as close to each other as we wish (in fact sometimes they can have the same position). Moreover, as we observe in section \ref{three}, if particles of a single type, say $i$, are far apart, then the monopole metric is likely to be exponentially close to yet another hyperk\"ahler metric. This metric is simpler than the monopole metric, but still given by transcendental functions. It is only when the particles of each type are far apart, that the asymptotic metric becomes algebraic and we are able to compute it explicitly.
\par
The proof uses the idea of our previous work \cite{BielCMP}, i.e. replacing solutions to Nahm's equations corresponding to monopoles with solutions defined on half-line. This gives a new moduli space whose metric is then computed by twistor methods. The main novelty is the way the metrics are compared. We prove (in Appendix B) a general theorem which allows us to deduce the estimates on all derivatives of components (in natural coordinates) of the difference of metric tensors from one-sided estimates on the metric tensors. Such a deduction is possible for two hyperk\"ahler metrics which are related via a complex-symplectic isomorphism providing that one of the metrics admits holomorphic coordinates in which the complex-symplectic form is standard and in which the components of the metric tensor are uniformly bounded.
\par
The paper is organized as follows. In the next section we collect some facts about $T^m$-invariant hyperk\"ahler metrics in dimension $4m$ of which our asymptotic metric is an example. In section \ref{one} we recall the description, due to Nahm \cite{Nahm} and to Hurtubise and Murray \cite{HurtMur}, of $SU(N)$-monopoles in terms of Nahm's equations. We also define there the moduli space of solutions to Nahm's equations whose metric will be the asymptotic metric. This moduli space is a hyperk\"ahler quotient of the product of several simpler moduli spaces, and in the following four sections we compute the metrics on these. In section \ref{topology} we discuss the topology of these moduli spaces. Finally, in section \ref{three}, we put the results together to obtain an explicit formula for the asymptotic metric (Theorem \ref{finalasymptoticmetric}). We also prove there that the rate of approximation is exponential (Theorem \ref{estimates}) and discuss the topology of the asymptotic moduli space.  
The appendix A deals with the question of identifying certain hyperk\"ahler quotients with corresponding complex-symplectic quotients which is needed in section \ref{topology}. In appendix B we prove the above-mentioned comparison theorem for Ricci-flat K\"ahler manifolds.

\section{Hyperk\"ahler quotients and $T^n$-invariant hyperk\"ahler metrics \label{zero}}

The Gibbons-Manton metric \cite{GM,GR} is an example of a $4m$-dimensional (pseudo)-hyperk\"ahler metric admitting a tri-Hamiltonian (hence isometric) action of the $n$-dimensional torus $T^m$.  Such metrics have particularly nice properties and were studied by several authors \cite{LR,HKLR,PP}. On the set where the action of $T^m$ is free such a metric can be locally written in the form:
\begin{equation} g=\Phi d\overline{\bf x}\cdot d\overline{\bf x}+\Phi^{-1}(d\overline{t}+A)^2,\label{torus-invariant}\end{equation}
where $\overline{\bf x}$ is the hyperk\"ahler moment map, $d\overline{t}$ is dual to the $(m\times 1)$-matrix of Killing  vector fields and the matrix $\Phi$ and the $1$-form $A$ depend only on the ${\bf x}_i$ and satisfy certain linear PDE's.  In particular, $\Phi$ determines the metric up to a gauge equivalence.
The set where the $T^m$-action is free can be viewed as a $T^m$-bundle over an open subset of ${\Bbb R}^{3m}$. For the Gibbons-Manton metric this open subset is the configuration space of $\tilde{C}_m({\Bbb R}^3)$ of $m$ distinct points in ${\Bbb R}^3$ (i.e. ${\Bbb R}^{3m}$ without the generalized diagonal) and 
\begin{equation}\Phi_{ij}=\begin{cases}-\frac{1}{\nu}-\sum_{k\neq i}\frac{1}{\|{\bf x}_i-{\bf x}_k\|} & \text{if $i=j$}\\ \frac{1}{\|{\bf x}_i-{\bf x}_j\|} & \text{if $i\neq j$}.\end{cases}\label{GM}\end{equation}
Here $\nu$ is the mass parameter. We can, in particular, take $\nu=\infty$ and $m=2$. Then the linearity of the equations for $\Phi$ and $A$ implies that, for any mapping $(i,j)\mapsto s_{ij}$ of $\{1,\ldots,m\}\times \{1,\ldots,m\}$ such that $s_{ij}=s_{ji}$ and $s_{ii}=0$ for $i,j=1,\ldots, m$, and for any constants $c_i$, $i=1,\dots,m$, the following matrix $\Phi$ defines a $T^m$-invariant (pseudo)-hyperk\"ahler metric:
\begin{equation}\Phi_{ij}=\begin{cases}c_i+\sum_{k\neq i}\frac{s_{ik}}{\|{\bf x}_i-{\bf x}_k\|} & \text{if $i=j$}\\ -\frac{s_{ij}}{\|{\bf x}_i-{\bf x}_j\|} & \text{if $i\neq j$}.\end{cases}\label{GMtype}\end{equation}
\par
The asymptotic metric on $M_{m_1,\dots,m_{N-1}}(\mu_1,\dots,\mu_N)$ will turn out to be of this form. Namely $m=m_1+\dots+m_{N-1}$ and, if we define the type $t(i)$ of an $i\leq m$ by $t(i)=\min\{k;i\leq \sum_{s\leq k}m_s\}$, then
\begin{equation} c_i=\mu_{k+1}-\mu_k\enskip \text{if $t(i)=k$}\label{c-asym}\end{equation}
and
\begin{equation}s_{ij}=\begin{cases}-2 & \text{ if $t(i)=t(j)$}\\
1 & \text{ if $|t(i)-t(j)|=1$}\\ 0 & \text{otherwise}.\end{cases}\label{s-asym} \end{equation}

\section{Moduli spaces of solutions to Nahm's equations \label{one}}

We shall define in this section several moduli spaces of solutions to Nahm's equations. All of these spaces carry hyperk\"ahler metrics. In particular, we shall recall, after Nahm \cite{Nahm} and Hurtubise and Murray \cite{HurtMur}, the description of the moduli spaces of $SU(N)$ monopoles with maximal symmetry breaking in terms of solutions to Nahm's equations. We shall also describe the asymptotic moduli spaces. We remark that from the point of view of hyperk\"ahler geometry many interesting metrics are obtained by replacing below the unitary group with an arbitrary compact Lie group. Such a generalization is straightforward, but, as our focus is on monopoles, we shall restrict ourselves to the unitary case.\bigskip

Nahm's equations are the following ODE's:
\begin{equation}\dot{T}_i+[T_0,T_i]+\frac{1}{2}\sum_{j,k=1,2,3}\epsilon_{ijk}[T_j,T_k]=0\;,\;\;\;\;i=1,2,3.\label{Nahm}\end{equation}
The functions $T_0,T_1,T_2,T_3$ are defined on some interval and are skew-hermitian and analytic. If the rank of the $T_i$ is $n$, then the space of solutions is acted upon by the gauge group ${\cal G}$ of $U(n)$-valued functions $g(t)$:
\begin{eqnarray} T_0&\mapsto & \Ad(g)T_0-\dot{g}g^{-1}\nonumber\\ T_i&\mapsto & \Ad(g)T_i\;,\;\;\qquad i=1,2,3.\label{action}\end{eqnarray}
We define fundamental moduli spaces of ${\frak u}(n)$-valued solutions $F_n(m;c)$ and $\tilde{F}_n(m;c)$. Here $m$ is a nonnegative integer  less than or equal to $n$ and $c$ is a positive real number ($c$ can be negative or zero for $\tilde{F}_n(m;c)$). The moduli spaces $F_n(m;c)$ correspond to monopoles with minimal symmetry breaking and are the basic building blocks from which all moduli spaces of framed $SU(N)$-monopoles with maximal symmetry breaking can be obtained by means of the hyperk\"ahler quotient construction. The spaces $\tilde{F}_n(m;c)$ play a similar role for the asymptotic metrics. They are defined as follows:
\begin{itemize}
\item Solutions in $F_n(m;c)$ are defined on $(0,c]$, while solutions in $\tilde{F}_n(m;c)$ are defined on $(0,\infty]$.
\item For a solution $(T_0,T_1,T_2,T_3)$ in either $F_n(m;c)$ or $\tilde{F}_n(m;c)$, $T_0$ and the $m\times m$ upper-diagonal blocks of $T_1,T_2,T_3$ are analytic at $t=0$, while the $(n-m)\times(n-m)$ lower-diagonal blocks have simple poles with residues defining the standard $(n-m)$-dimensional irreducible representation of ${\frak su}(2)$. The off-diagonal blocks are of the form $t^{(n-m-1)/2}\times(\text{\it analytic in $t$})$.
\item A solution in $F_n(m;c)$ is analytic at $t=c$, while a solution in $\tilde{F}_n(m;c)$ approaches a diagonal limit at $+\infty$ exponentially fast. Furthermore $(T_1(+\infty),T_2(+\infty),T_3(+\infty))$ is a regular triple, i.e. its centralizer consists of diagonal matrices.
\item The gauge group for $F_n(m;c)$ consists of gauge transformations $g$ with $g(0)=g(c)=1$, while the gauge group for $\tilde{F}_n(m;c)$ has the Lie algebra consisting of functions $\rho:[0,+\infty)\rightarrow {\frak u}(n)$  such that
\begin{itemize}
\item[(i)] $\rho(0)=0$ and $\dot{\rho}$ has a diagonal limit at $+\infty$;
\item[(ii)] $(\dot{\rho}-\dot{\rho}(+\infty))$ and $[\tau,\rho]$ decay exponentially fast for any regular diagonal matrix $\tau\in{\frak u}(n)$;
\item[(iii)] $c\dot{\rho}(+\infty)+\lim_{t\rightarrow +\infty} (\rho(t)-t\dot{\rho}(+\infty))=0$.
\end{itemize}
\end{itemize}
\begin{remark} Alternately, the space  $\tilde{F}_n(m;c)$ can be viewed as the moduli space of solutions defined on $[-c,+\infty]$ with the gauge group given by the transformation which are exponentially close to $\exp(ht)$ for some diagonal $h$.\label{shift}\end{remark}
The tangent space at a solution $(T_0,T_1,T_2,T_3)$ can be identified, for both $F_n(m;c)$ and $\tilde{F}_n(m;c)$, with the space of solutions to the following system of linear equations:
\begin{equation}\begin{array}{c} \dot{t}_0+[T_0,t_0]+[T_1,t_1]+[T_2,t_2]+[T_3,t_3]=0,\\ 
\dot{t}_1+[T_0,t_1]-[T_1,t_0]+[T_2,t_3]-[T_3,t_2]=0,\\
\dot{t}_2+[T_0,t_2]-[T_1,t_3]-[T_2,t_0]+[T_3,t_1]=0,\\  
\dot{t}_3+[T_0,t_3]+[T_1,t_2]-[T_2,t_1]-[T_3,t_0]=0.\end{array}\label{tangent}\end{equation}
$F_n(m;c)$ carries a hyperk\"ahler metric is defined by
\begin{equation}\|(t_0,t_1,t_2,t_3)\|^2=\int_{0}^c\sum_0^3\|t_i\|^2 \label{metric},\end{equation}
while $\tilde{F}_n(m;c)$ possesses an indefinite (and possibly degenerate) hyperk\"ahler metric given by:
\begin{equation}\|(t_0,t_1,t_2,t_3)\|^2=c\sum_0^3\|t_i(+\infty)\|^2+\int_0^{+\infty}\sum_0^3\left(\|t_i(s)\|^2-\|t_i(+\infty)\|^2\right)ds. \label{smetric}\end{equation}
The moduli space $F_n(m;c)$ has a tri-Hamiltonian action of $U(n)\times U(m)$ given by gauge transformations $g$ with arbitrary values at $t=c$ and with $g(0)$ being block-diagonal with the off-diagonal blocks equal to $0$ and  the $(n-m)\times(n-m)$ lower-diagonal block being identity. Both $U(n)$ and $U(m)$ act freely. The hyperk\"ahler moment map for the action of $U(n)$ is $(-T_1(c),-T_2(c),-T_3(c))$, while the one for the action of $U(m)$ is $\pi(T_1(0),T_2(0),T_3(0))$, where $\pi$ denotes the projection onto the $m\times m$ upper-diagonal block.
\par  
The moduli space $\tilde{F}_n(m;c)$ has a similarly defined free tri-Hamiltonian action of $U(m)$. In addition, it has a free tri-Hamiltonian action of the diagonal torus $T^n\leq U(n)$ given by gauge transformations which are asymptotic to $\exp(-th+\lambda h)$ for a diagonal $h$ and real $\lambda$. The moment map for this action is $(T_1(+\infty),T_2(+\infty),T_3(+\infty))$.\bigskip

We shall now consider hyperk\"ahler quotients of various products of ${F}_n(m;c)$ and $\tilde{F}_n(m;c)$. We observe that the hyperk\"ahler quotient construction of say, ${F}_n(m;c)\times {F}_n(l;c^\prime)$ matches solutions $(T_0(t),T_1(t),T_2(t),T_3(t))$ in ${F}_n(m;c)$ with $(-T_0(c+c^\prime-t),-T_1(c+c^\prime-t),-T_2(c+c^\prime-t),-T_3(c+c^\prime-t))$   for a $(T_0(t),T_1(t),T_2(t),T_3(t))$ in $\tilde{F}_n(m;c)$. The resulting space can be identified with the moduli space of solutions to Nahm's equations on $[0,c+c^\prime]$ having appropriate poles at $t=0$ and at $t=c+c^\prime$. We recall that the triple backslash denotes hyperk\"ahler quotient (in all our constructions the moment map is canonical and we quotient its $0$-set). We have basic hyperk\"ahler isomorphisms:
\begin{eqnarray} \bigl({F}_n(m;c)\times {F}_n(n;c^\prime)\bigr)/\!/\!/U(n) & \simeq & {F}_n(m;c+c^\prime)\nonumber\\  
\bigl({F}_n(m;c)\times \tilde{F}_n(n;c^\prime)\bigr)/\!/\!/U(n) & \simeq & \tilde{F}_n(m;c+c^\prime).\label{quotient}\end{eqnarray}
The group acts diagonally on the product. In particular all $\tilde{F}_n(m;c)$ can be obtained from the $\tilde{F}_n(n;c)$ and the ${F}_n(m;c)$.
\par
We now define auxiliary moduli spaces $F_{n,m}(c,c^\prime)$, $F_{\tilde{n},m}(c,c^\prime)$, $F_{n,\tilde{m}}(c,c^\prime)$, $F_{\tilde{n},\tilde{m}}(c,c^\prime)$. Here $n,m$ are arbitrary positive integers and $c,c^\prime$ are arbitrary positive real numbers. The spaces are defined as follows:
\begin{itemize} 
\item if $n<m$, then $F_{n,m}(c,c^\prime)=\bigl(F_n(n;c)\times F_m(n;c^\prime)\bigr)/\!/\!/U(n)$. The spaces with a tilde over $n$ or $m$ are obtained by replacing the corresponding $F$ with $\tilde{F}$;
\item if $n>m$, then $F_{n,m}(c,c^\prime)=\bigl(F_n(m;c)\times F_m(m;c^\prime)\bigr)/\!/\!/U(m)$ and similarly for the other spaces;
\item if $n=m$, then $F_{n,n}(c,c^\prime)$ is the hyperk\"ahler quotient of $F_n(n;c)\times F_n(n;c^\prime)\times {\Bbb H}^n$ by the diagonal action of $U(n)$.
\end{itemize}
\begin{remark}Thus these moduli spaces  consist of $\frak{u}(m)$-valued solutions $T_i^-$ on $[-x,0)$  and of $\frak{u}(n)$-valued  solutions  $T_i^+$ on $(0,y]$, where $x=c^\prime$ or $-\infty$ and $y=c$ or $y=+\infty$, with matching conditions at $t=0$: if $n>m$ (resp. $n<m$), then the limit  of the $m\times m$ upper-diagonal block of $T_i^+$ (resp. $T_i^-$)  at $t=0$ is equal to the limit of $T_i^-$ (resp. $T_i^+$); if $n=m$, then there exists a vector $(V,W)\in {\Bbb C}^{2n}$ such that $(T_2^++iT_3^+)(0_+)-(T_2^-+iT_3^-)(0_-)= VW^T$ and $T_1^+(0_+)-T_1^-(0_-)=(|V|^2-|W|^2)/2$. The gauge transformations $g(t)$ satisfy similar matching conditions: if $n\neq m$, then the upper-diagonal $m\times m$ block is continuous, the lower-diagonal block is identity at $t=0$ and the off-diagonal blocks vanish to order $(n-m-1)/2$; if $n=m$, then  $g(t)$ is continuous at $t=0$.\label{match}\end{remark}

Notice that $F_{n,m}(c,c^\prime)$ is isomorphic, as a hyperk\"ahler manifold, to $F_{m,n}(c^\prime,c)$ and similarly for $F_{\tilde{n},\tilde{m}}(c,c^\prime)$.
We can now define the moduli spaces we are really interested in. Let us fix an integer $N$ and consider functions $\sigma:\{1,\ldots,N-1\}\rightarrow {\Bbb N}\sqcup {\Bbb N}$ is arbitrary and $\mu:\{1,\ldots,N\}\rightarrow {\Bbb R}$ is increasing. We shall denote the second copy of ${\Bbb N}$ by $\tilde{\Bbb N}$ and write its elements as $\tilde{1},\tilde{2},\tilde{3},\ldots$. We define the moduli space $F_\sigma(\mu)$ as a hyperk\"ahler quotient of $F_{\sigma_1}(c_1)\times F_{\sigma_2}(c_2,c_2^\prime)\times\ldots\times F_{\sigma_{N-1}}(c_{N-1},c_{N-1}^\prime)\times F_{\sigma_{N}}(c_{N}^\prime)$. Here $c_i+c_{i+1}^\prime=\mu(i+1)-\mu(i)$, $\sigma_1,\sigma_N\in{\Bbb N}\sqcup {\Bbb N}$ and $\sigma_i:\{1,2\}\rightarrow {\Bbb N}\sqcup {\Bbb N}$ for $2\leq i\leq N-1$. Furthermore $\sigma_1=\sigma(1)$, $\sigma_i(1)=\sigma(i)$, $\sigma_i(2)=\sigma(i-1)$ for $2\leq i\leq N-1$, and $\sigma_N=\sigma(N-1)$. Finally $F_n(c),F_{\tilde{n}}(c)$ denote $F_n(0;c)$ and $\tilde{F}_n(0;c)$. The group by which we quotient is a product of unitary groups and of tori acting on this product space: we take the diagonal action of $U(n)$ on $F_{\sigma_i}(c_i,c^\prime_i)\times F_{\sigma_{i+1}}(c_{i+1},c^\prime_{i+1})$ (or $F_{\sigma_1}(c_1)\times F_{\sigma_2}(c_2,c_2^\prime)$ for $i=1$ and similarly for $i=N-1$)  if $\sigma(i)=n$ and  the diagonal action of $T^n$ if $\sigma(i)=\tilde{n}$.
\begin{remark}The moduli space $F_\sigma(\mu)$ should be viewed as consisting of solutions to Nahm's equations on $N-1$ ``intervals" $I_i$, $i=1,\dots,N_1$ with matching conditions at the boundary points. If $\sigma(i)\in {\Bbb N}$, then $I_i=[\mu(i),\mu(i+1)]$, while if $\sigma(i)\in \tilde{\Bbb N}$, then $I_i=[\mu(i),+\infty)\cup(-\infty,\mu(i+1)]$ (see Remark \ref{shift}). The solutions satisfy matching conditions of Remark \ref{match} at each $\mu_i$ and are continuous at each infinity. The gauge transformations satisfy matching conditions of Remark \ref{match} at each $\mu_i$ and are exponentially close to $\exp(h_it+p_i)$ near $\pm \infty$ in $I_i$, $\sigma(i)\in \tilde{\Bbb N}$, for some diagonal matrices $h_i,p_i$.\label{tilde(M)}\end{remark}

A theorem of Hurtubise and Murray \cite{HurtMur}, giving a full proof of the correspondence found by Nahm \cite{Nahm}, and  generalizing the $SU(2)$ case due to Hitchin \cite{Hit1} can be phrased as follows (this formulation uses the  connectivity of the moduli space of $SU(N)$-monopoles due to Jarvis \cite{Jarvis}):
\begin{theorem} The moduli space $M_{m_1,\ldots,m_{N-1}}(\mu_1,\ldots,\mu_N)$ of framed $SU(N)$ \linebreak monopoles of charge $(m_1,\ldots,m_{N-1})$ and the symmetry breaking at infinity equal to $(\mu_1,\ldots,\mu_N)$, $\mu_i$ distinct, is diffeomorphic to the moduli space $F_\sigma(\mu)$ with $\sigma(i)=m_i$ and $\mu(i)=\mu_{i}$. \qed \label{mon-Nahm} \end{theorem}

\begin{remark}It is expected, but at present not known (except for $N=2$ \cite{Nak}), that this diffeomorphism is an isometry. Nevertheless, the twistorial character of Hurtubise and Murray's construction shows that this diffeomorphism preserves the three complex structures and, hence, the Levi-Civita connection. Thus, the geodesics are the same. \label{isometry}\end{remark}

Our aim is to show that the metric on $F_\sigma(\mu)$ with $\sigma(i)=m_i$ and $\mu(i)=\mu_{i}$ is asymptotic to the metric on $F_{\tilde{\sigma}}(\mu)$ with $\tilde{\sigma}(i)=\tilde{m}_i$ and $\mu(i)=\mu_i$. We shall first compute the metric on $F_{\tilde{\sigma}}(\mu)$.

\section{Complex structures on $F_n(m;c)$ and $\tilde{F}_n(n;c)$\label{two}}

All moduli spaces described in the previous section have an isometric action of $SU(2)$ or $SO(3)$ rotating the complex structures and therefore all complex structures are equivalent. We shall consider the complex structure $I$ and describe the complex coordinates (and the complex symplectic form $\omega_2+i\omega_3$) on $F_n(m;c)$ and $\tilde{F}_n(n;c)$. All other moduli spaces can be described as open subsets of complex-symplectic quotients of products of these.\\ 
We set $\alpha=T_0+iT_1$ and $\beta=T_2+iT_3$. The Nahm equations can be then written as one complex and one real equation:
\begin{eqnarray} & &\frac{d\beta}{dt} = [\beta,\alpha]\label{complex}\\
 & &\frac{d\,}{dt}(\alpha+\alpha^\ast) =[\alpha^\ast,\alpha]+[\beta^\ast,\beta].\label{real}\end{eqnarray}
First, we consider $F_n(m;c)$ (cf. \cite{Hurt, BielCMP}).\\
Let $E_1,\ldots,E_n$ denote the standard basis of ${\Bbb C}^n$. There is a unique solution $w_1$ of the equation
\begin{equation} \frac{dw}{dt}=-\alpha w\label{alphato0}\end{equation}
with
\begin{equation} \lim_{t\rightarrow 0}\left(t^{-(n-m-1)/2}w_1(t)-E_{m+1}\right)=0\label{w1}.\end{equation}
Setting $w_i(t)=\beta^{i-1}(t)w_1(t)$, we obtain a solution to \eqref{alphato0} with 
$$ \lim_{t\rightarrow 0}\left(t^{i-(n-m+1)/2}w_i(t)-E_{m+i}\right)=0.$$
In addition there are solutions $u_1,\ldots,u_m$ to \eqref{alphato0} whose last $n-m$ components vanish to order $(n-m+1)/2$, and which are linearly independent at $t=0$.
The complex gauge transformation $g(t)$ with $g^{-1}=(u_1,\ldots,u_m,w_1,\ldots,w_{n-m})$ makes $\alpha$ identically zero and sends $\beta(t)$ to the constant matrix (cf. \cite{Hurt})

\begin{equation}B=\left(\begin{array}{ccc|cccc} &  &  & 0 &\ldots & 0 & g_1\\
& h & & \vdots & & \vdots & \vdots\\ &  &  & 0 &\ldots & 0 & g_m \\ \hline 
f_1 & \ldots & f_m & 0 &\ldots & 0 & e_1\\
0 &\ldots & 0 & 1 & \ddots & & e_2\\
\vdots & & \vdots & & \ddots & \ddots & \vdots \\0 &\ldots & 0 & 0&\ldots & 1 & e_{n-m} \end{array}\right).\label{betaconst}\end{equation}

The mapping $(\alpha,\beta)\rightarrow (g(c),B)$ gives a biholomorphism between $(F_n(m;c),I)$ and $Gl(n,{\Bbb C})\times{\frak gl}(m,{\Bbb C})\times {\Bbb C}^{n+m}$ \cite{BielJLMS}. The action of $Gl(n,{\Bbb C})$ is given by the right translations, and the action of $Gl(m,{\Bbb C})$ is given by $p\cdot\bigl(h,f,g,e,g(c)\bigr)=(php^{-1},fp^{-1},pg,e,pg(c))$, where for the last term we embedded $Gl(m,{\Bbb C})$ in $Gl(n,{\Bbb C})$ as the $m\times m$ upper-diagonal block.  We can compute the complex symplectic form $\omega=\omega_2+i\omega_3$. We denote by $b,\hat{b}$ vectors tangent to the space of $B$'s in \eqref{betaconst} and by $\rho,\hat{\rho}$ right-invariant vector fields on $Gl(n,{\Bbb C})$. We have \cite{BielJLMS}:
\begin{equation}\omega\left((\rho,b),(\hat{\rho},\hat{b})\right)=\tr\bigl(\rho\hat{b}-\hat{\rho}b-B[\rho,\hat{\rho}]\bigr).\label{form}\end{equation}
Now we consider the complex structure of $\tilde{F}_n(n;c)$. Let ${\frak n}$ be a unipotent algebra corresponding to the Cartan algebra of diagonal matrices. We consider the open dense subset $\tilde{F}({\frak n})$ of $\tilde{F}_n(n;c)$ defined as  the set of all solutions $(\alpha,\beta)=(T_0+iT_1,T_2+iT_3)$  such that the intersection of the sum of positive eigenvalues of $\text{ad}(iT_1(+\infty))$ with with the centralizer $C(\beta(+\infty))$ is contained in ${\frak n}$. We observe that, since $(T_1(+\infty),T_2(+\infty),T_3(+\infty))$ is a regular triple, the projection of $T_1(+\infty)$ onto $C(\beta(+\infty))$ is a regular element. Now, as in \cite{BielCMP}, we use results of Biquard \cite{Biq} to deduce that $\tilde{F}({\frak n})$ is biholomorphic to an open subset of $Gl(n,{\Bbb C})\times_N {\frak b}$ where $N=\exp{\frak n}$ and ${\frak b}={\frak d}+{\frak n}$, ${\frak d}$ denoting the diagonal matrices. Briefly, the element $g$ of $Gl(n,{\Bbb C})$ is given by the value at $t=0$ of the complex gauge transformation $g(t)$ which makes $(0,\beta(+\infty)+n)$ into $(\alpha,\beta)$.\\ The charts $\tilde{F}({\frak n})$ are glued as follows: $[g,d+n] \sim [g^\prime,d^\prime+n^\prime]$ if and only if  $n\in {\frak n},n^\prime \in {\frak n}^\prime$,  and either ${\frak n}^\prime\subset {\frak n}$ and there exists an $m\in N$ such that $gm^{-1}=g^\prime,\Ad(m)(d+{\frak n})=d^\prime+{\frak n}^\prime$ or vice versa (i.e. ${\frak n}\subset {\frak n}^\prime$ etc.). 
\par
We remark that $F_{\emptyset}$ is an open dense subset biholomorphic to an open subset of $Gl(n,{\Bbb C})\times {\frak d}$. We shall denote this subset by $\tilde{F}_n^{\rm reg}(n;c)$. If $b_d,\hat{b}_d$ denote vectors tangent to the space of diagonal matrices and $\rho,\hat{\rho}$ denote this time left-invariant vector fields on $Gl(n,{\Bbb C})$, then the form $\omega$ is given by \cite{BielCMP}:
\begin{equation}\omega=-\tr\bigl(b_d\hat{\rho}-\rho\hat{b}_d-[\rho,\hat{\rho}]\beta_d\bigr).\label{form2}\end{equation}

All other moduli spaces $\tilde{F}_n(m;c)$ and $F_\sigma(\mu)$ can be viewed as hyperk\"ahler quotients of products of ${F}_n(m;c)$ and $\tilde{F}_n(n;c)$. Thus, as complex-symplectic manifolds, they are isomorphic to open subsets of complex-symplectic quotients of the corresponding complex-symplectic manifolds computed above. The description of the latter quotients is straightforward. Let us remark that Hurtubise showed in \cite{Hurt} that if $\sigma(i)=m_i$, $i=1,\ldots,N-1$, then $F_\sigma(\mu)$ (i.e. the moduli space of $SU(N)$-monopoles) is biholomorphic to the space of based rational maps from ${\Bbb C}P^1$ to $SU(N)/T$ (maximal torus) of degree $(m_1,\ldots,m_{N_1})$.

\section{Complex-symplectic structure of $F_{\tilde{n},\tilde{m}}(c,c^\prime)$, $n>m$ \label{four}}

Our aim is to calculate the metric on $F_\sigma(\mu)$ where $\sigma(i)=\tilde{m}_i$. This space has dimension $4p=4(m_1+\ldots+m_{N-1})$ and admits a tri-Hamiltonian action of $T^p$. By the definition of $F_\sigma(\mu)$, it is a hyperk\"ahler quotient, by a torus, of the product spaces $\tilde{F}_n(0;c)$ and $F_{\tilde{n},\tilde{m}}(c,c^\prime)$. The metric on $\tilde{F}_n(0;c)$ was calculated in \cite{BielCMP} - it is the Gibbons-Manton metric \cite{GM} with the mass parameter $-1/c$. It remains to calculate the metric on $F_{\tilde{n},\tilde{m}}(c,c^\prime)$. For convenience we shall write
$$\tilde{F}_{n,m}(c,c^\prime):=F_{\tilde{n},\tilde{m}}(c,c^\prime).$$ 
The dimension of this space is $4(n+m)$ and it has a tri-Hamiltonian action of an $(n+m)$-dimensional torus. 
\par
The space $\tilde{F}_{n,m}(c,c^\prime)$ should be thought as consisting of solutions to Nahm's equations on $(-\infty,0)\cup (0,\infty)$, which are ${\frak u}(m)$-valued on $(-\infty,0)$,  ${\frak u}(n)$-valued on $(0,\infty)$, and satisfy appropriate matching conditions at zero. In what follows we shall usually say ``$\tilde{F}_{n,m}(c,c^\prime)$ is biholomorphic to \dots" rather than ``$\tilde{F}_{n,m}(c,c^\prime)$ is biholomorphic to an open subset of \dots". This never leads to any problems.

We consider the space $\tilde{F}_{n,m}(c,c^\prime)$ for $n>m$. From its description as a complex-symplectic quotient, $\tilde{F}_{n,m}(c,c^\prime)$ is given by charts of the form $\{(b_-,(g,b_+)\}\in  {\frak b}\times  \bigl(Gl(n,{\Bbb C})\times_{N^\prime} {\frak b}^\prime\bigr)$ such that $gb_+g^{-1}$ is of the form \eqref{betaconst} with $h=b_-$. Let us consider the chart on which ${\frak b}={\frak d}_m$ and  ${\frak b}^\prime={\frak d}_n$ (${\frak d}_m$ and ${\frak d}_n$ denote $m\times m$ and $n\times n$ diagonal matrices). Let us write the elements of ${\frak d}_m$ as $\diag(\kappa_1,\ldots,\kappa_m)$ and the elements of ${\frak d}_n$ as $\diag(\beta_1,\ldots,\beta_n)$. Let $q_+(z)=\prod (z-\beta_i)$ and $q_-(z)=\prod (z-\kappa_i)$. We assume that the roots of both these polynomials are distinct and we consider multiplication by $z$ on ${\Bbb C}[z]/(q_+)$. It is a linear operator which, in the basis 
\begin{equation}\frac{\prod_{j\neq i}(z-\beta_j)}{\prod_{j\neq i}(\beta_i-\beta_j)},\qquad {i=1,\ldots,n},\label{basis1}\end{equation} 
is the diagonal matrix $\diag(\beta_1,\ldots,\beta_n)$. On the other hand, in the basis   \begin{equation}\frac{\prod_{j\neq i}(z-\kappa_j)}{\prod_{j\neq i}(\kappa_i-\kappa_j)},q_-(z),\ldots,z^{n-m-1}q_-(z),\qquad {i=1,\ldots,m},\label{basis2}\end{equation}
 the multiplication by $z$ is given by a matrix of the form \eqref{betaconst} with $f_i=1\big/\prod_{j\neq i}(\kappa_i-\kappa_j)$. Let $Z$ be the matrix transforming the basis \eqref{basis1} into \eqref{basis2}. Then any $g$ which sends $\diag(\beta_1,\ldots,\beta_n)$ to a matrix of the form \eqref{betaconst} can be written as
\begin{equation}\diag(v_1^{-1},\ldots,v_m^{-1},1,\ldots,1)Z\diag(u_1,\ldots,u_n).\label{g-Z}\end{equation}
We shall now compute $Z$. We introduce one more basis of ${\Bbb C}[z]/(q_+)$:
\begin{equation} 1,\ldots,z^{n-1}.\label{basis3}\end{equation}
The passage from \eqref{basis1} to \eqref{basis3} is given by $V(\beta_1,\ldots,\beta_n)^{-1}$, where $V(\beta_1,\ldots,\beta_n)$ is the Vandermonde matrix, i.e. its $(i,j)$-th entry is $(\beta_i)^{j-1}$. We then compute the passage from \eqref{basis3} to \eqref{basis2} as given by the matrix
\begin{equation} L=\left(\begin{array}{c|c} V_{\kappa} & W\\ \hline 0 & H \end{array}\right),\label{el}\end{equation}
where $V_{\kappa}=V(\kappa_1,\ldots,\kappa_m)$, $W_{ij}=(\kappa_i)^{m+j-1}$ and $H$ is upper-triangular with $H_{ij}=(-1)^{i-j}H_{i-j}$, where $H_k$ denotes the $k$-th complete symmetric polynomial in $\kappa_1,\ldots,\kappa_m$, i.e. the sum of all monomials of degree $k$.
\begin{remark} The factorization $Z=LV^{-1}$ is unique only if $\beta_i\neq\kappa_j$ for all $i,j$. If, for instance, $\beta_1=\kappa_1$, then the above $g$ sends  $\diag(\beta_1,\ldots,\beta_n)$ to a matrix of the form \eqref{betaconst} with $g_1=0$. However, there is then another $g$ which makes $f_1=0$.\label{unique}\end{remark}

We calculate the complex symplectic form on $\tilde{F}_{n,m}(c,c^\prime)$. The chart where $\beta_i\neq \beta_j$, $\kappa_r\neq\kappa_s$, $\beta_i\neq \kappa_s$ for all $i,j=1,\dots,n$, $i\neq j$, $r,s=1,\dots,m$, $r\neq s$, can be described as consisting of pairs $\bigl((g_{-},\kappa_d),(g_{+},\beta_d)\bigr)$, where $\kappa_d=\diag(\kappa_1,\dots,\kappa_m)$, $\beta_d=\diag(\beta_1,\dots,\beta_n)$, $g_{-}=V_\kappa^{-1}\diag(v_1,\dots,v_m)$, $g_{+}=V_\kappa^{-1}L V_\beta^{-1}\diag(u_1,\dots,u_n)$. According to the formula \eqref{form2}, the complex symplectic form in this chart is equal to: 
$$\omega=-\tr\bigl(k_d\tilde{\rho}_--\rho_-\tilde{k}_d-\kappa_d[\rho_-,\tilde{\rho}_-]+b_d\tilde{\rho}_+-\rho_+\tilde{b}_d-\beta_d[\rho_+,\tilde{\rho}_+]\bigr).$$
Here $k_d,\rho_-,b_d,\rho_+$ are dual to, respectively, $\kappa_d,(g_-)^{-1}dg_-,d\beta_d, (g_+)^{-1}dg_+$.\newline
The first three terms can be computed as in \cite{BielCMP} and give
\begin{equation}\sum_{i=1}^m \frac{dv_i}{v_i}\wedge  d\kappa_i-\sum_{i<j}\frac{d\kappa_i\wedge d\kappa_j}{\kappa_i-\kappa_j} .\label{omega-}\end{equation}
To compute the remaining two terms let us write $X=V_{\kappa}^{-1}L$ and $Y=V_{\beta}^{-1}\diag(u_1,\ldots,u_n)$. Let us also write $\beta_c=Y\beta_d Y^{-1}$ and $b_c,x,y$ for vector fields dual to $\beta_c,X^{-1}dX, Y^{-1}dY$. Then $\rho_+=Y^{-1}xY+y$ and $b_c=Yb_dY^{-1}+Y[y,\beta_d]Y^{-1}$. Thus the last three terms in the above formula can be rewritten as
\begin{multline*}-\tr\bigl(b_d\tilde{\rho}_+-\rho_+\tilde{b}_d-\beta_d[\rho_+,\tilde{\rho}_+]\bigr)= -\tr\Bigl(b_d\tilde{y}-y\tilde{b}_d +b_dY^{-1}\tilde{x}Y- Y^{-1}xY \tilde{b}_d \\ -\beta_d[y,\tilde{y}] -\beta_d[y,Y^{-1}\tilde{x}Y]-\beta_d[Y^{-1}xY,\tilde{y}] -Y\beta_dY^{-1}[x,\tilde{x}]\Bigr)=\omega_+ -\tr\Bigl(Yb_dY^{-1}\tilde{x}- xY \tilde{b}_dY^{-1} \\ -Y\beta_dY^{-1}[YyY^{-1},\tilde{x}]-Y\beta_dY^{-1}[x,Y\tilde{y}Y^{-1}] -Y\beta_dY^{-1}[x,\tilde{x}]\Bigr)=\omega_+-\tr\bigl(b_c\tilde{x}-x\tilde{b}_c- \beta_c[x,\tilde{x}]\bigr).\end{multline*}
Here $\omega_+=-\tr\bigl(b_d\tilde{y}-y\tilde{b}_d-\beta_d[y,\tilde{y}]\bigr)$, which, again as in \cite{BielCMP}, is equal to
\begin{equation}\sum_{i=1}^n \frac{du_i}{u_i}\wedge  d\beta_i-\sum_{i<j}\frac{d\beta_i\wedge d\beta_j}{\beta_i-\beta_j} .\label{omega+}\end{equation}
For the remaining terms we observe that $d\beta_c$ is upper triangular and $X^{-1}dX$ is strictly upper-triangular. Hence the remaining terms vanish and the complex-symplectic form on $\tilde{F}_{n,m}(c,c^\prime)$ is given by: 
\begin{equation}\omega= \sum_{i=1}^n \frac{du_i}{u_i}\wedge  d\beta_i-\sum_{i<j}\frac{d\beta_i\wedge d\beta_j}{\beta_i-\beta_j} +\sum_{i=1}^m \frac{dv_i}{v_i}\wedge d\kappa_i-\sum_{i<j}\frac{d\kappa_i\wedge d\kappa_j}{\kappa_i-\kappa_j}.\label{omega} \end{equation}

\section{Twistor space and the metric of $\tilde{F}_{n,m}(c,c^\prime)$, $n>m$\label{five}}

First, we compute the twistor space of $F_n(m;c)$. Let $\zeta$ be the affine coordinate on ${\Bbb C}P^1$. The twistor space of any hyperk\"ahler manifold admitting an action of $SU(2)$ rotating the complex structures can be trivialized using just two charts $\zeta\neq \infty$ and $\zeta\neq 0$.
For a moduli space of solutions to Nahm's equations this is achieved by putting  $\eta=\beta+(\alpha+\alpha^\ast)\zeta-\beta^\ast\zeta^2$, $u=\alpha-\beta^\ast\zeta$ over $\zeta\neq\infty$  and $\tilde{\eta}=\beta/\zeta^{2}+(\alpha+\alpha^\ast)/\zeta-\beta^\ast$, $\tilde{u}=-\alpha^\ast-\beta/\zeta$ over $\zeta\neq 0$. Then, over $\zeta\neq 0,\infty$, we have $\tilde{\eta}=\eta/\zeta^2$, $\tilde{u}=u-\eta/\zeta$. Moreover, the real structure is $\zeta\mapsto -1/\bar{\zeta}$, $\eta\mapsto -\eta^\ast/\bar{\zeta}^2$, $u\mapsto -u^\ast+\eta^\ast/\bar{\zeta}$ (cf. \cite{Dan2,Biq2}).
\par

We consider the matrix \eqref{betaconst}.
We have
\begin{lemma} The $(i,j)$-th entry of $\exp{Bt/\zeta}$ is of the form $\frac{1}{(i-j)!}(t/\zeta)^{i-j}+O(t^{i-j+1})$ for $i>j>m$.\end{lemma}
\begin{pf} We write $(Bt/\zeta)^k$ in the block form as 
$$\begin{pmatrix} P(k) & Q(k)\\ R(k) & S(k) \end{pmatrix}. $$ 
One then checks by induction that:
$$ S(k)_{(r,s)}=\begin{cases} 0 & \text{if $k<r-s$}\\
(t/\zeta)^{r-s} & \text{if $k=r-s$}\\ O(t^{r-s+1}) & \text{if $k>r-s$}\end{cases}$$
and similarly for the other blocks.
\end{pf}

Therefore the $(m+1)$-th column, denoted by $\tilde{p}_{m+1}$ of $g^{-1}(t)\exp\{Bt/\zeta\}$ is of the form $t^{n-m-1}p +O(t^{n-m})$, for some vector $p$. This means that $p$ belongs to the $-(n-m-1)/2$-eigenspace of $\Res\tilde{u}$, and so is of the form $aE_n$ ($E_n$ is the $n$-th vector of the standard basis), for some constant $a$. Computing the $t^{n-m-1}$-term of the last entry of $\tilde{p}_{m+1}$ gives $(\Res \eta)^{n-m-1}E_{m+1}\bigl(\zeta^{n-m-1}(n-m-1)!\bigr)^{-1}=\zeta^{-(n-m-1)}$, and so $a=\zeta^{-(n-m-1)}$. Thus, as in \eqref{w1}, $\tilde{w}_1= \zeta^{n-m-1}p_{m+1}$ is a solution $\tilde{w}_1(t)$ to $\frac{d}{dt}\tilde{w}_1=-\tilde{u}\tilde{w}_1$ with
$$\lim_{t\rightarrow 0}\bigl(t^{-(n-m-1)/2}w_1(t)-E_n\bigr)=0.$$
In the same vein we see that $\tilde{w}_i(t)=\zeta^{n-m+1-2i}\tilde{p}_{m+i}$, where $\tilde{p}_{m+i}$ is the $(m+i)$-th column of $g^{-1}(t)\exp\{Bt/\zeta\}$. In other words $\tilde{g}(t)=d(\zeta)\exp\{Bt/\zeta\}g(t)$ where $$d(\zeta)=\diag\{1,\ldots,1,\zeta^{-(n-m-1)},\ldots,\zeta^{(n-m-1)}\}.$$
\par
Similar computations show that the real structure sends $B$ to $-r(\zeta)\bigl(B^\ast/\bar{\zeta}^2\bigr)r(\zeta)^{-1}$ and $g$ to $r(\zeta)\exp\{B^\ast/\bar{\zeta}\}(g^\ast)^{-1}$ where
\begin{equation}r_{ij}(\zeta)=\begin{cases} 0 & \text{if $i+j\neq n+m+1$}\\
 (-1)^{j-1}\bar{\zeta}^{n+m+1-2j} & \text{if $i+j= n+m+1$}.\end{cases} \label{rij}\end{equation}

Now we consider the subset of $\tilde{F}_n(n;c)$, where the eigenvalues of $\beta(+\infty)$ are distinct. We have assigned to each element of this set the pair $(\beta(+\infty);g)$. We know that $\tilde{\beta}(+\infty)=\beta(+\infty)/\zeta^2$. The argument in section 2 shows then that $\tilde{g}=g\exp\{-c\beta(+\infty)/\zeta\}$. The real structure sends $g$ to $(g^\ast)^{-1}\exp\{c\beta(+\infty)^\ast/\bar{\zeta}\}$.

Now we wish to calculate the twistor space of $\tilde{F}_{n,m}(c,c^\prime)$ for $n>m$.
This space is a hyperk\"ahler quotient of $F_n(m;1)\times\tilde{F}_n(m;c-1)\times\tilde{F}_m(m;c^\prime)$.
On the subset corresponding to the same chart as in section \ref{four} (i.e. ${\frak b}={\frak d}_m$ and ${\frak b}^\prime={\frak d}_n$), the coordinates are given by $\beta(-\infty),\beta(+\infty)$ and $g$  such that $g\beta(+\infty)g^{-1}$ is of the form \eqref{betaconst} with $h=\beta(-\infty)$. This $g$ can be written as $(g_3)^{-1}g_1g_2$, where $g_1$ (resp. $g_2$, resp. $g_3$) is the $g$ considered above for $F_n(m;1)$ (resp. $\tilde{F}_n(m;c-1)$, resp. $\tilde{F}_m(m;c^\prime)$). It follows that $$\tilde{g}=\exp\{c^\prime\beta(-\infty)/\zeta\}(g_3)^{-1}d(\zeta)\exp\{B/\zeta\} g_1 g_2\exp\{-(c-1)\beta(+\infty)/\zeta\}$$
which can be rewritten as
$$ \exp\{c^\prime\beta(-\infty)/\zeta\}d(\zeta) (g_3)^{-1}g_1g_2 \exp\{-c\beta(+\infty)/\zeta\}.$$

Therefore 
$$ \tilde{g}=\exp\{c^\prime\beta(-\infty)/\zeta\}d(\zeta) g \exp\{-c\beta(+\infty)/\zeta\}.$$

We now compute the twistor space in coordinates $\kappa_1,\ldots,\kappa_m$,$v_1,\ldots,v_m$,$\beta_1,\ldots,\beta_n$, $u_1,\ldots,u_n$ of section \ref{four}. We know that $\tilde{\kappa}_i=\kappa_i/\zeta^2$ and $\tilde{\beta}_j=\beta_j/\zeta^2$. The matrix $g$ is given by \eqref{g-Z}, where $Z=LV^{-1}$ where $L$ is described by \eqref{el} and $V$ is the Vandermonde matrix for $\beta_1,\ldots,\beta_n$. We obtain equations (here $v_i=\tilde{v}_i=1$ and $\kappa_i=\tilde{\kappa}_i=0$ for $i>m$):
$$\tilde{v}_i^{-1}\tilde{u}_j\tilde{Z}_{ij}=\exp\{(c^\prime \kappa_i-c\beta_j)/\zeta\} d_i(\zeta) Z_{ij}v_i^{-1}u_j.$$
In addition $\tilde{Z}_{ij}=Z_{ij}$ if $i\leq m$ and $\tilde{Z}_{ij}=\zeta^{2i-2}Z_{ij}$ if $i>m$. Hence
$$\tilde{v}_i^{-1}\tilde{u}_j=\begin{cases}\exp\{(c^\prime \kappa_i-c\beta_j)/\zeta\}v_i^{-1}u_j & \text{if $i\leq m$}\\ \zeta^{-n-m+1}\exp\{-c\beta_j/\zeta\}v_i^{-1}u_j & \text{if $i>m$}. \end{cases}$$
As $v_i=\tilde{v}_i=1$ for $i>m$, we finally obtain
\begin{xalignat}{1} & \tilde{v}_i=\zeta^{-n-m+1} \exp\{-c^\prime \kappa_i/\zeta\}v_i\\
& \tilde{u}_j=\zeta^{-n-m+1}\exp\{-c\beta_j/\zeta\}u_j.\end{xalignat}
Finally, the real structure is computed as in \cite{BielCMP}:
\begin{xalignat}{2}  \beta_i\mapsto -\bar{\beta}_i/\bar{\zeta}^2, & & 
u_i\mapsto \bar{u}_i^{-1}(1/\bar{\zeta})^{n+m-1}e^{c\bar{\beta}_i/\bar{\zeta}} \prod_{j\neq i}(\bar{\beta}_i-\bar{\beta}_j)\prod_{j=1}^m(\bar{\beta}_i-\bar{\kappa}_j) 
&\label{real1} \\
 \kappa_i\mapsto -\bar{\kappa}_i/\bar{\zeta}^2, & & v_i\mapsto \bar{v}_i^{-1}(1/\bar{\zeta})^{n+m-1}e^{c^\prime\bar{\kappa}_i/\bar{\zeta}} \prod_{j\neq i}(\bar{\kappa}_i-\bar{\kappa}_j)\prod_{j=1}^n(\bar{\kappa}_i-\bar{\beta}_j)
&\label{real2} \end{xalignat}
We now have to calculate the real sections. First of all we have
\begin{xalignat}{2} & \beta_i(\zeta)=z_i+2x_i\zeta-\bar{z}_i\zeta^2 & & \text{for $i=1,\ldots,n,$}\\
&  \kappa_i(\zeta)=z_{n+i}+2x_{n+i}\zeta-\bar{z}_{n+i}\zeta^2 & & \text{for $i=1,\ldots,m,$}\end{xalignat}
where $p_i=(z_i,x_i)\in {\Bbb C}\times{\Bbb R}$ are such that $p_i\neq p_j$ if $i\neq j$. These curves of genus $0$ should be thought of as spectral curves of individual monopoles. Let $S_i$ denote either $\beta_i$ or $\kappa_i$. Two curves $S_i$ and $S_j$  intersect in a pair of distinct points $a_{ij}$ and $a_{ji}$, where
\begin{equation} a_{ij}=\frac{(x_i-x_j)+r_{ij}}{\bar{z}_i-\bar{z}_j}, \quad r_{ij}=\sqrt{(x_i-x_j)^2 +|z_i-z_j|^2}.\label{aij}\end{equation}
As in \cite{BielCMP}, if $i,j\leq n$, then $u_i$ has a zero at $a_{ji}$ and is nonzero at $a_{ij}$. Similarly, if $i,j> n$, then $v_{i-n}$ has a zero at $a_{ji}$ and is nonzero at $a_{ij}$. Let us consider what happens when $i\leq n$ and $j>n$ (and no other curves intersect $S_i$ at $a_{ji}$). First of all, computing the characteristic polynomial of \eqref{betaconst} gives \cite{Hurt}:
\begin{equation}\det (\eta -B)=\det(\eta-h)(\eta^{n-m}-e_{n-m}\eta^{n-m-1}-\ldots -e_1) -f(\eta-h)_{\rm adj}g,\label{det}\end{equation}
from which we conclude that $f_{j-n}g_{j-n}$ is zero at both $a_{ij}$ and $a_{ji}$. This implies, since $f_{i}=v_i/\prod_{s\neq i}(\kappa_i-\kappa_s)$, that $v_{j-n}$ is zero precisely when $f_{j-n}$ is zero. Now, if the passage from $\diag(\beta_1,\ldots,\beta_n)$ to \eqref{betaconst} is given by the matrix $G$ of the form \eqref{g-Z} with $Z=LV^{-1}$, then $G_{j-n,s}=0$ if $s\neq i$ and $G_{j-n,i}=v_{j-n}^{-1}u_i$. This has two implications: 1) $u_i$ is zero if and only if $v_{j-n}$ is, and  2) $g_j=0$ in \eqref{betaconst}.  Hence, in this situation, $v_{j-n}\neq 0$. Thus $u_i$ and $v_{j-n}$ are zero at exactly one 
of the two points of intersection of $S_i$ and $S_j$. Furthermore, since $\kappa_k(\zeta)$ does not intersect $\beta_l(\zeta)$ at $a_{ij}$ or $a_{ji}$ if $k\neq j-n$ or $l\neq i$, we conclude that $f_kg_k\neq 0$ at $a_{ij}$ or $a_{ji}$ if $k\neq j-n$. Thus $v_k(a_{ij})\neq 0$ and $v_k(a_{ji})\neq 0$ for $k\neq j-n$. Since $G(a_{ij})$ and $G(a_{ji})$ are invertible we also have $u_l(a_{ij})\neq 0$ and $u_l(a_{ji})\neq 0$ for $l\neq i$.
\par
Summing up, $u_i(\zeta)$, $i\leq n$, and $v_i(\zeta)$, $i\leq m$, are of the form:
\begin{equation}
u_i(\zeta)=A_i\prod\begin{Sb}j\leq n\\ j\neq i\end{Sb}(\zeta-a_{ji}) \prod_{j>n}(\zeta-c_{ij})e^{c(x_i-\bar{z}_i\zeta)},\label{ui}\end{equation}
\begin{equation}
v_i(\zeta)=B_i\prod\begin{Sb}j\leq m\\j\neq i\end{Sb}(\zeta-a_{j+n,i+n}) \prod_{j\leq n} (\zeta-c_{i+n,j})e^{c^\prime(x_i-\bar{z}_i\zeta)}.\label{vi}\end{equation}
Here $c_{ij}$ can be either $a_{ij}$ or $a_{ji}$ and is at present undetermined.
The reality condition implies that
$$A_i\bar{A}_i=\prod\begin{Sb}j\leq n\\ j\neq i\end{Sb}(x_i-x_j+r_{ij}) \prod_{j>n}\bigl(\pm(x_i-x_j)+r_{ij}\bigr),$$
$$B_i\bar{B}_i=\prod\begin{Sb}j\leq m\\j\neq i\end{Sb}(x_{i+n}-x_{j+n}+r_{i+n,j+n}) \prod_{j\leq n} \bigl(\pm(x_{i+n}-x_j)+r_{i+n,j}\bigr),$$
where the undetermined signs are positive if $c_{ij}=a_{ji}$ and negative if $c_{ij}=a_{ij}$. By continuity, these formulae extend to the case when more than two $S_i$ intersect at a point.
\par
We can now compute the metric on $\tilde{F}_{n,m}(c,c^\prime)$ up to the above sign indeterminacy. This metric is of the form \eqref{torus-invariant} and it is enough to compute the matrix $\Phi$. The complex symplectic form on the twistor space is given by the formula \eqref{omega} and it follows from it that each factor in \eqref{ui} (resp. \eqref{vi}) together with the corresponding factor of $|A_i|$ (resp. $|B_i|$) gives a separate contribution to $\Phi$ (the coefficient of $\zeta$ in the expansion of $\omega$ is the K\"ahler form $\omega_1$). The factors of $u_i$, indexed by $j\leq n$, together with the exponential term, describe exactly the twistor space of the Gibbons-Manton metric \eqref{GM} with mass parameter $-1/c$, as computed in Proposition 6.2 of \cite{BielCMP}, providing that the complex symplectic form is taken to be the first two terms of \eqref{omega}. Thus these terms contribute
$$c-\sum\begin{Sb}j\leq n\\ j\neq i\end{Sb}\frac{1}{r_{ij}}$$
to $\Phi_{ii}$, $i\leq n$, and $1/r_{ij}$ to $\Phi_{ij}$ for $i,j\leq n$, $i\neq j$. An exactly parallel statement holds for the factors of $v_i$ indexed by $j\leq m$ plus the exponential term. The remaining factors contribute terms in $\frac{du_{i}}{u_{i}}\wedge d\beta_i$ and $\frac{dv_{i}}{v_{i}}\wedge d\kappa_i$. The calculation in the proof of Theorem 6.4 in \cite{BielCMP} shows that the contribution of these factors to the matrix $\Phi$ are Gibbons-Manton-like terms with positive or negative sign. Thus we conclude that the matrix $\Phi$ for $\tilde{F}_{n,m}(c,c^\prime)$ is given by \eqref{GMtype} with 
\begin{equation} c_i=\begin{cases} c & \text{if $i\leq n$}\\ c^\prime & \text{if $i> n$},\end{cases}\label{ci}\end{equation} 
\begin{equation} s_{ij}=\begin{cases} -1 & \text{if $i,j\leq n$ or $i,j>n$}\\
(-1)^{\epsilon_{ij}} & \text{if $i\leq n, j>n$ or $i>n,j\leq n$}.\end{cases}\label{sij}\end{equation} 
Here $\epsilon_{ij}=0$ if $c_{ij}=a_{ij}$ and $\epsilon_{ij}=1$ if $c_{ij}=a_{ji}$. We shall eventually see (Lemma \ref{epsilonij}) that all $\epsilon_{ij}$ are equal to zero.

\section{ The metric on $\tilde{F}_{n,n}(c,c^\prime)$ \label{six}} 

This space is the hyperk\"ahler quotient of $\tilde{F}_n(n;c)\times \tilde{F}_n(n;c^\prime)\times {\Bbb H}^n$ by the diagonal action of $U(n)$.
According to section \ref{two} its complex charts can be described as $\bigl(b_-,(g,b_+), (V,W)\bigr)\in  {\frak b}\times  \bigl(Gl(n,{\Bbb C})\times_{N^\prime} {\frak b}^\prime\bigr)\times{\Bbb C}^{2n}$
with $gb_+g^{-1}=b_-+VW^T$.
\par
Our first step is to calculate the complex-symplectic form on the set where ${\frak b}={\frak b}^\prime={\frak d}$. Let us write $\beta_d^+=\diag(\beta_1^+,\dots,\beta_n^+)$ for $b_+$ and $\beta_d^-=\diag(\beta_1^-,\dots,\beta_n^-)$ for $b_-$ on this set. 
The choice of our chart implies that $\beta_i^+\neq \beta_j^+$ and $\beta_i^-\neq \beta_j^-$ for $i\neq j$. In addition we suppose that $\beta_i^-\neq \beta_j^+$ for all $i,j\leq n$. Then one of $\beta_d^-,\beta_d^+$, say $\beta_d^-$, is invertible. Since all components of $W$ must be nonzero  (otherwise the spectra of $\beta_d^-$ and $\beta_d^+$  are not disjoint), $W$ is cyclic for $\beta_d^-$. Consider the basis given by columns of $\bigl((\beta_d^-)^{n-1}W,(\beta_d^-)^{n-2}W,\ldots,W\bigr)^T$, in which $\beta^-$ is  of form \eqref{betaconst} (with $m=0$) and $W^T=(0,0,\ldots,1)$. Thus, $\beta^+$ is also of form \eqref{betaconst}. We can therefore describe this chart as consisting of pairs $\bigl((g_-,\beta_d^-),(g_+,\beta_d^+)\bigr)$ with $g_-\beta_d^-(g_-)^{-1}$ and $g_+\beta_d^+(g_+)^{-1}$ both of form \eqref{betaconst} (with $m=0$). We have $g_{\pm}=V(\beta_1^{\pm},\ldots, \beta_n^{\pm})^{-1}\diag (u_1^\pm,\ldots,u_n^\pm)$. The form $\omega$, via the complex-symplectic quotient, can be written as ($b_d^\pm,\rho_{\pm}$ are dual to $\beta_d^\pm, (g_\pm)^{-1}dg_\pm$):
$$\omega=-\tr\bigl(b_d^+\tilde{\rho}_+-\rho_+\tilde{b}_d^+-\beta_d^+[\rho_+, \tilde{\rho}_+]+b_d^-\tilde{\rho}_--\rho_-\tilde{b}_d^--\beta_d^-[\rho_-, \tilde{\rho}_-]\bigr),$$
which  can be computed as in \cite{BielCMP} giving:
\begin{equation}\omega= \sum_{i=1}^n \frac{du_i^-}{u_i^-}\wedge  d\beta_i^- -\sum_{i<j}\frac{d\beta_i^-\wedge d\beta_j^-}{\beta_i^--\beta_j^-} + \sum_{i=1}^n \frac{du_i^+}{u_i^+}\wedge  d\beta_i^+ -\sum_{i<j}\frac{d\beta_i^+\wedge d\beta_j^+}{\beta_i^+-\beta_j^+}.\label{omega+-}\end{equation}

We now wish to compute the twistor space of $\tilde{F}_{n,n}(c,c^\prime)$. 
 We proceed as in the previous section. A calculation done there shows that the coordinates $g,\beta_d^-,\beta_d^+$ (here $g\beta_d^+ g^{-1}=\beta_d^-+VW^T$) change from $\zeta\neq \infty$ to $\zeta\neq 0$ as:
$$\tilde{\beta}_d^\pm=\beta_d^\pm/\zeta^2,\qquad \tilde{g}=\exp\{c^\prime\beta_d^-\}g\exp\{-c\beta_d^+\}. $$
Moreover, since the twistor space of ${\Bbb H}^n$ is simply $O(1)\otimes {\Bbb C}^{2n}$, we have
$$\tilde{V}=V/\zeta,\qquad \tilde{W}=W/\zeta.$$
We wish to pass to coordinates $\beta_i^\pm,u_i^\pm$, $i=1,\dots,n$. The passage from the basis in which $\beta^\pm$ are diagonal to the one in which they are of the form \eqref{betaconst} is achieved by the matrix 
$$H=\bigl((\beta_d^-)^{n-1}W,(\beta_d^-)^{n-2}W,\dots,W\bigr)^T.$$
Thus $g_-=Hd$ and $g_+=Hdg$ for some diagonal $d$. We have  $\tilde{d}=d\exp\{-c^\prime\beta_d^-\}$ ($d$ is $g_-$ in the chart in which $b_-$ is diagonal). On the other hand we have written $g_{\pm}=V(\beta_1^{\pm},\ldots, \beta_n^{\pm})^{-1}\diag (u_1^\pm,\ldots,u_n^\pm)$. Comparing the two formulae in the two charts, we conclude that $u^\pm_i(\zeta)$ changes from $\zeta\neq \infty$ to $\zeta\neq 0$ as:
$$ \tilde{u}_i^+=\zeta^{1-2n}\exp\{-c\beta^+_i/\zeta\}u^+_i,\quad 
 i=1,\dots,n,$$
$$ \tilde{u}_i^-=\zeta^{1-2n}\exp\{-c^\prime\beta^-_i/\zeta\}u^-_i,\quad 
 i=1,\dots,n.$$
The real structure is given by the formulae \eqref{real1} and \eqref{real2} with $m=n$, $\beta_i=\beta_i^+,\kappa_i=\beta_i^-,u_i=u_i^+,v_i=u_i^-$.

 We have to know what happens to the $u^\pm_i(\zeta)$ when two curves $\beta_i^\pm(\zeta)$ intersect. As in the previous section we write $S_i(\zeta)=\beta^+_i(\zeta)=z_i+2x_i\zeta -\bar{z}_i\zeta^2$, $S_{n+i}(\zeta)=\beta^-_i(\zeta)=z_{n+i}+2x_{n+i}\zeta -\bar{z}_{n+i}\zeta^2$, and we denote the intersection points of $S_i$ and $S_j$ by $a_{ij},a_{ji}$. These are given by the formula \eqref{aij}. 
\par
Consider first the intersection of $S_i$ and $S_j$ where both $i$ and $j$ are either less than or equal to $n$ (and no other $S_k$ intersect at $a_{ij},a_{ji}$). We can still assume generically that the spectra of $\beta_d^-$ and $\beta_d^+$ are disjoint and, hence, $W$ is cyclic for $\beta_d^-$. Then $H$ is invertible at $a_{ij},a_{ji}$ and so are $g_{\pm}$. We compute, as in \cite{BielCMP}, that each $u_i^\pm$ has a zero at the intersection point $a_{ji}$ and is nonzero at $a_{ij}$, and all other $u_s^\pm$, $s\neq i,j$ are nonzero at both   $a_{ij}$ and $a_{ji}$. The same argument works in the case when both $i$ and $j$ are greater than $n$.
\par
Now consider the intersection of $S_i$ and $S_j$ where $i\leq n$ and $j>n$. In the chart in which $\beta^-$ is diagonal, we had $g\beta_d^+g^{-1}=\beta_d^-+VW^T$. Thus $\det(\beta_d^- - \beta_i^+1+VW^T)=0$. Since $VW^T$ has rank one, and so all its $k\times k$, $k>1$, minors vanish, we have for any diagonal matrix $d=\diag(d_1,\dots,d_n)$ the formula
\begin{equation}\det(d+VW^T)=\prod_k d_k +\sum_{k}\left(V_kW_k\prod_{l\neq k} d_l\right).\label{det2}\end{equation}
In our case $d_j=0$ and $d_k\neq 0$ for $k\neq j$. We conclude that $V_jW_j$ vanish at both $a_{ij}$ and $a_{ji}$. However both $V_j$ and $W_j$ are sections of $O(1)$ and so have exactly one zero. Thus $W_j$ vanishes at either $a_{ij}$ or $a_{ji}$ (and only one of them). Furthermore, if we consider the diagonal matrix $d=\beta_d^- - \beta_s^-1$, $s\neq j$, then, by the above argument, the non-vanishing of
$\det(d+VW^T)$ implies that $V_sW_s$ does not vanish at either $a_{ij}$ or $a_{ji}$ if $s\neq j$. In summary, it is precisely the $j$-th column of $H$ that vanishes at either $a_{ij}$ or $a_{ji}$. Thus the same statement holds for both $g_-$ and $g_+$, and, as $g_{\pm}=V(\beta_1^{\pm},\ldots, \beta_n^{\pm})^{-1}\diag (u_1^\pm,\ldots,u_n^\pm)$, we conclude that both $u_j^-$ and $u_j^+$ vanish at either $a_{ij}$ or $a_{ji}$ and no other $u_s^\pm$ vanishes at either $a_{ij}$ or $a_{ji}$. This means that $u_i^+$ is given by the formula \eqref{ui} and $u_i^-$ is given by the formula \eqref{vi} (with $n=m$). Once more, the formulae extend to the non-generic case. The remainder of the previous section can be now repeated word by word, and we conclude that the metric on $\tilde{F}_{n,n}(c,c^\prime)$ is of the form \eqref{torus-invariant} with $\Phi$ given by \eqref{GMtype} where the $c_i$ and $s_{ij}$ are given by \eqref{ci} and \eqref{sij}.

\section{Topology of $\tilde{F}_{n,m}(c,c^\prime)$ \label{topology}}

We shall discuss the topology of $\tilde{F}_{n,m}(c,c^\prime)$. This space can be viewed as a moduli space of solutions to Nahm's equations defined on $(-\infty,0]\cup(0,+\infty)$ with the appropriate matching at $0$. The tri-Hamiltonian action of $T^{n+m}=T^n\times T^m$ gives us the moment map to ${\Bbb R}^3\otimes {\Bbb R}^{n+m}$ which is simply $$\Bigl(\bigl(T_1(+\infty),-T_1(-\infty)\bigr), \bigl(T_2(+\infty),-T_2(-\infty)\bigr),\bigl(T_2(+\infty),-T_2(-\infty)\bigr)\Bigr).$$
Before stating the result let us recall that a basis of the second homology $H_2\bigl(\tilde{C}_{p}({\Bbb R}^3),{\Bbb Z}\bigr)$ of a configuration space $\tilde{C}_{p}({\Bbb R}^3)$ is given by the $p(p-1)/2$ $2$-spheres
\begin{equation}S_{ij}^2=\{({\bf x}_1,\dots,{\bf x}_p)\in{\Bbb R}^3\otimes {\Bbb R}^p; |{\bf x}_i-{\bf x}_j|={\it const},\enskip {\bf x}_k={\it const}\enskip\text{if}\enskip k\neq i,j\}\label{basis}\end{equation}
 where $i<j$. We have:
\begin{proposition} The above moment map induces a homeomorphism between the orbit space of $\tilde{F}_{n,m}(c,c^\prime)$ and  $\tilde{C}_{n}({\Bbb R}^3)\times \tilde{C}_{m}({\Bbb R}^3)$. The  set of principal $T^{n+m}$-orbits of $\tilde{F}_{n,m}(c,c^\prime)$ maps to  $\tilde{C}_{n+m}({\Bbb R}^3)$ and as a $T^{n+m}$-bundle is determined by the element $(h_1,\ldots,h_{n+m})$ of $H^2\bigl(\tilde{C}_{n+m}({\Bbb R}^3),{\Bbb Z}^{n+m}\bigr)$ given by $$h_k(S_{ij}^2)=\begin{cases} s_{ij} &\text{if $k=i$}\\-s_{ij} &\text{if $k=j$}\\0 &\text{otherwise,}\end{cases}$$ where the $s_{ij}$ are given by \eqref{sij}.\label{bundle}\end{proposition}
\begin{pf} 
 Let us fix an element $(\tau^+,\tau^-)$ of $ \tilde{C}_{n}({\Bbb R}^3)\times \tilde{C}_{m}({\Bbb R}^3)$. Identify $\tau^+$ with a regular triple $(\tau^+_1,\tau^+_2,\tau_3^+)$ of diagonal $n\times n$ matrices and similarly $\tau^-$ with a regular triple $(\tau^-_1,\tau^-_2,\tau_3^-)$ of diagonal $m\times m$ matrices.  As in Proposition 5.2 of \cite{BielCMP} the space of $T^{n+m}$-orbits mapping to $(\tau^+,\tau^-)$ can be identified with the set of solutions to Nahm's equations with $T_0\equiv 0$ and having values conjugate to $(\tau^+,-\tau^-)$ at $+\infty$ and at $-\infty$. If $n>m$, this space is diffeomorphic to the hyperk\"ahler quotient $X$ of the product $M(\tau^+_1,\tau^+_2,\tau_3^+)\times F_n(m;1)\times M(\tau^-_1,\tau^-_2,\tau_3^-)$ by $U(n)\times U(m)$, where the $M$'s are Kronheimer's hyperk\"ahler structures on $Gl(n,{\Bbb C})/(T^n)^{\Bbb C}$ and $Gl(m,{\Bbb C})/(T^m)^{\Bbb C}$ \cite{Kron}. If $n=m$, this space is diffeomorphic to the hyperk\"ahler quotient $X$ of the product $M(\tau^+_1,\tau^+_2,\tau_3^+)\times {\Bbb H}^n\times M(\tau^-_1,\tau^-_2,\tau_3^-)$ by $U(n)$. 
\par
The first statement will be proved if we can show that these hyperk\"ahler quotients are single points. First we show that the corresponding complex-symplectic quotient, with respect to a generic complex structure $I$ (i.e. one in which $M(\tau^\pm_1,\tau^\pm_2,\tau_3^\pm)$ are biholomorphic to regular adjoint orbits), are single points.\newline
{\bf (1) $n>m$.} Let $M(\tau^+_1,\tau^+_2,\tau_3^+)$ be complex-symplectic isomorphic to the adjoint orbit $O^+$ of $\diag(\beta_1^+,\dots,\beta_n^+)$ ($\beta^+_i$ distinct) and $M(\tau^-_1,\tau^-_2,\tau_3^-)$ to the adjoint orbit $O^-$ of $\diag(\beta_1^-,\dots,\beta_m^-)$ ($\beta^-_i$ distinct). First of all, the complex symplectic quotient of $M(\tau^+_1,\tau^+_2,\tau_3^+)\times F_n(m;1)$ by $Gl(n,{\Bbb C})$ can be identified with the set $U$ of elements of $O^+$ which are of the form \eqref{betaconst}. Then the zero-set of the complex moment map for the action of $Gl(m,{\Bbb C})$ on $U\times O^-$ can be identified with the set $Y$ of matrices of the form  \eqref{betaconst} which belong to $O^+$ and such that $h$ belongs to $O^-$. We have to show that $Y$ is a single orbit of $Gl(m,{\Bbb C})$. Since the $\beta_i^-$ are distinct we can diagonalize $h$. Then the equation \eqref{det} shows that the $e_i$'s and the products $f_ig_i$ are determined. Thus we obtain a single $({\Bbb C}^\ast)^m$-orbit.\newline 
{\bf (2) $n=m$.} We make the same assumption about the complex-symplectic structure of the two $M$'s. The zero set of the complex moment map for the action of $Gl(n,{\Bbb C})$ on $O^+\times O^-\times {\Bbb C}^{2n}$ is the set $\{(a,b,V,W)\in O^+\times O^-\times {\Bbb C}^{n} \times {\Bbb C}^{n};\, a=b+VW^T\}$. Again we have to show that this set is a single orbit of $Gl(n,{\Bbb C})$. Let us diagonalize $b$ and use the formula \eqref{det2} with $d=b-\eta 1$. Substituting $\beta_i^-$ for $\eta$ shows that $V_iW_i$ is determined, $i=1,\dots,n$. We obtain a single $({\Bbb C}^\ast)^n$-orbit.
\par
We remark that the above proof shows that the action of $G^{\Bbb C}$, where  $G^{\Bbb C}$ is $Gl(n,{\Bbb C})\times Gl(m,{\Bbb C})$  in case (1) or $Gl(n,{\Bbb C})$  in case (2),  on the zero-set of the complex moment map has closed orbits of the form $G^{\Bbb C}/T^{\Bbb C}$ for some subtorus  $T$ of $G$.
\par
Thus, to prove the first statement, we have to show that the complex-symplectic and the hyperk\"ahler quotient coincide. The proof of this requires a substantial detour from the main line of argument and will be given in the appendix A. Let us remark that Hurtubise's argument \cite{Hurt} for matching solutions to Nahm's equations on two (or more) intervals cannot be adapted to the case of two half-lines (in this case his Lemma 2.19 will not provide any information).
\par
It is clear from the description of the sections of the twistor space - formulae \eqref{ui} and \eqref{vi} - that the action is free precisely over $\tilde{C}_{n+m}({\Bbb R}^3)$. To determine the principal bundle, one merely has to repeat the calculation in the proof of Proposition 6.3 in \cite{BielCMP}.\end{pf}

\begin{corollary} The action of $T^{n+m}$ on $\tilde{F}_{n,m}(c,c^\prime)$ extends to the global action of $({\Bbb C}^\ast)^{n+m}$ with respect to any complex structure.\label{C*}\end{corollary}
\begin{pf} This is equivalent to showing that, if we fix $\zeta\in {\Bbb C}P^1$, then the $u_i(\zeta)$ and $v_j(\zeta)$ of \eqref{ui} and \eqref{vi} can take arbitrary complex values (with appropriate degenerations at the intersection points of the $\beta_i(\zeta)$). If, for example $\zeta=0$, then the $z_i$ are fixed and one solves for the $x_i$. One shows that a solution always exists and by the previous result the corresponding point lies in  $\tilde{F}_{n,m}(c,c^\prime)$. \end{pf}

\section{Asymptotic comparison of metrics \label{three}}

We consider the moduli space $M_{m_1,\ldots,m_{N-1}}(\mu_1,\ldots,\mu_N)$ of $SU(N)$ monopoles with maximal symmetry breaking. We wish to compare the metric on $F_{m_1,\ldots,m_{N-1}}(\mu_1,\ldots,\mu_N)$ (whose Levi-Civita connection coincides with that on $M_{m_1,\ldots,m_{N-1}}(\mu_1,\ldots,\mu_N)$) with the metric on $F_{\tilde{m}_1,\ldots,\tilde{m}_{N-1}}(\mu_1,\ldots,\mu_N)$. As discussed in Remark \ref{tilde(M)}, this space consists of solutions to Nahm's equations on the union of $I_k$ where $I_k=[\mu_k,+\infty)\cup (-\infty,\mu_{k+1}]$ with matching conditions at the endpoints of each $I_k$. It will be convenient to write 
$$I_k=[[\mu_k,\mu_{k+1}]]$$
and denote the ``middle point" $\pm \infty$ by $\infty_k$.  We shall also use double brackets for any connected subset of $[[\mu_k,\mu_{k+1}]]$.
\par
The space $F_{\tilde{m}_1,\ldots,\tilde{m}_{N-1}}(\mu_1,\ldots,\mu_N)$ should be thought of as consisting of $m=m_1+\dots +m_{N-1}$ particles with phases. The positions of particles are ${\bf x}_i^k=(x_i^k,\text{Re}\,z_i^k,\text{Im}\,z_i^k)$, $i\leq m_k$, $k=1,\dots N-1$, where $\diag( x^k_1,\dots,x^k_{m_k})=\sqrt{-1}T_1(\infty_k)$ and  $\diag( z^k_1,\dots,z^k_{m_k})=(T_2+\sqrt{-1}T_3)(\infty_k)$. We put
\begin{equation} R_k=\min \{|{\bf x}_i^k-{\bf x}_j^k|; i\neq j\}.\label{R}\end{equation}
Let us also write
\begin{equation} Z_k=\min \{|z_i^k-z_j^k|; i\neq j\},\label{Z}\end{equation}
and denote by $F_{\tilde{m}_1,\ldots,\tilde{m}_{N-1}}^{\rm reg}(\mu_1,\ldots,\mu_N)$ the subset of $F_{\tilde{m}_1,\ldots,\tilde{m}_{N-1}}(\mu_1,\ldots,\mu_N)$ where $Z_k>0$ for $k=1,\dots,N-1$. This subset depends on the chosen complex structure (which is $I$ in the case at hand).
If we write for this  complex structure, as in section \ref{two}, $\alpha$ for $T_0+iT_1$, $\beta$ for $T_2+iT_3$, then we can define the subset $F_{m_1,\ldots,m_{N-1}}^{\rm reg}(\mu_1,\ldots,\mu_N)$ of $F_{m_1,\ldots,m_{N-1}}(\mu_1,\ldots,\mu_N)$ as the set of $(\alpha,\beta)$ such that the eigenvalues of $\beta$ restricted to the $k$-th interval, $k=1,\dots,N-1$, are distinct. 
\par
We define subset $U(\gamma,\delta,C)$ of $F_{\tilde{m}_1,\ldots,\tilde{m}_{N-1}}(\mu_1,\ldots,\mu_N)$ as follows
\begin{equation} U(\gamma,\delta,C)=\{{\bf x}; \min_k Z_k({\bf x})\geq \delta,\enskip \min_k R_k({\bf x})\geq C,\enskip \zeta^T\Phi\zeta\geq \gamma|\zeta|^2\enskip \forall\zeta\in{\Bbb R}^m\},\label{Udelta}\end{equation}
where $\Phi$ is given by \eqref{GMtype}-\eqref{s-asym} and $m=\sum m_k$.
\par
We have canonical local complex coordinates on  $F_{\tilde{m}_1,\ldots,\tilde{m}_{N-1}}^{\rm reg}(\mu_1,\ldots,\mu_N)$:
\begin{equation} (w_1,\dots,w_m):=\{z_i^k,u_i^k; \;i=1,\dots,m_k,\enskip k=1,\dots, N-1\}\label{cancoor}\end{equation}
where the $u_i^k$ are given by the local ${\Bbb C}^m$-action, $m=\sum m_k$.
\par
Let $\tilde{g},g$ denote the metrics on  $F_{\tilde{m}_1,\ldots,\tilde{m}_{N-1}}^{\rm reg}(\mu_1,\ldots,\mu_N)$ and  $F_{m_1,\ldots,m_{N-1}}^{\rm reg}(\mu_1,\ldots,\mu_N)$ respectively, and let $\Sigma$ be the product of symmetric groups $\prod_{k=1}^{N-1}\Sigma_{m_{k}}$. 
We can now state the two main results of the paper.
\begin{theorem} The hyperk\"ahler metric on $F_{\tilde{m}_1,\dots,\tilde{m}_{N-1}}(\mu_1,\dots,\mu_N)$ is determined by the matrix $\Phi$ of the form \eqref{GMtype} with the $c_i$ and $s_{ij}$ given by \eqref{c-asym} and \eqref{s-asym}.\label{finalasymptoticmetric}\end{theorem}
\begin{theorem} There exists a complex-symplectic isomorphism $\phi$ from \linebreak $F_{\tilde{m}_1,\ldots,\tilde{m}_{N-1}}^{\rm reg}(\mu_1,\ldots,\mu_N)/\Sigma$ to  $F_{m_1,\ldots,m_{N-1}}^{\rm reg}(\mu_1,\ldots,\mu_N)$ with the following property:
\par
 Let us write $$\phi^\ast g-\tilde{g}=\real\sum S_{ij}dw_i\otimes d\bar{w_j}$$
in coordinates \eqref{cancoor}. Then, for any positive $\gamma,\delta$, there is a $C=C(\gamma,\delta)$ such that on the set $U(\gamma,\delta,C)$ defined by \eqref{Udelta},
we have
\begin{equation} |D^l S_{ij}|\leq A_l e^{-\lambda R},\quad l=0,1,2,\dots,\label{DkSij}\end{equation}
where  $R=\min\{R_k; k=1,\dots,N-1\}$ and $A_l,\lambda>0$ are constants depending only on $\gamma,\delta$.\label{estimates}\end{theorem} 
\begin{remarks} 1. For a possible generalization see the discussion at the end of the section.\newline
2. One can alternately use the coordinates given by positions and phases of particles and obtain a completely analogous statement. This follows at once from the explicit formulae for the metric and the twistor space of $F_{\tilde{m}_1,\ldots,\tilde{m}_{N-1}}(\mu_1,\ldots,\mu_N)$. We see that the coordinate change map and its inverse have all derivatives uniformly bounded on $U(\gamma,\delta,C)$ (cf. \cite{Besse}, section 13.F for the case of Taub-NUT).    \end{remarks}
The proof of Theorems \ref{finalasymptoticmetric} and \ref{estimates} will be separated into several parts. We shall write $M$ for $F_{m_1,\ldots,m_{N-1}}^{\rm reg}(\mu_1,\ldots,\mu_N)$ and $\tilde{M}$ for $F_{\tilde{m}_1,\ldots,\tilde{m}_{N-1}}^{\rm reg}(\mu_1,\ldots,\mu_N)$. As usual, we use the same letter to denote constants varying from line to line.\newline
{\bf Part 1: Construction of $\phi$.}
This is completely analogous to the $SU(2)$ case. 
One goes via an intermediate moduli space $M_I$ consisting of solutions $(\alpha,\beta)$ to the complex Nahm equation which are constant and diagonal on each $[[\mu_k+c,\mu_{k+1}-c]]$ for some $c<\min\{(\mu_{k+1}-\mu_k)/2;k=1,\dots,N-1\}$ and satisfy appropriate matching conditions at each $\mu_k$, modulo gauge transformations $g(t)$ which satisfy the matching conditions of Remark \ref{tilde(M)}. In particular $g(t)=  \exp\{h_kt-p_k\}$ near $\infty_k$ for some complex diagonal matrices  
for diagonal $h_k,p_k$. The passage from $M_I$ to $M$ is given by restricting these solutions to the union of $[[\mu_k,(\mu_k+\mu_{k+1})/2]]\cup [[(\mu_k+\mu_{k+1})/2, \mu_{k+1}]]$, viewing them as solutions to the complex Nahm equation on $[\mu_1,\mu_{N-1}]$ and solving the real equation as in \cite{Hurt}.
\par
The map from $M$ to $M_I/\Sigma$ is defined by using a complex gauge transformation to make an element $(\alpha,\beta)$ of $M$ constant and diagonal on each $[\mu_k+c,\mu_{k+1}-c]$, cut off at the center of each interval and extend trivially onto $[[\mu_k,\mu_{k+1}]]$.
\par
The passage from $\tilde{M}$ to $M_I$ is given, as in section \ref{two}, by making a solution constant and diagonal on each $[[\mu_k+c,\mu_{k+1}-c]]$ by a complex gauge transformation with $g(\infty_k)=1$ and $g(t)=1$ on each $[\mu_k,\mu_k+c/2]\cup[\mu_{k+1}-c/2,\mu_{k+1}]$. The inverse mapping is given by first solving the real Nahm equation by a complex transformation which is exponentially close to $\exp\{h_kt-p_k\}$ near $\infty_k$ for some diagonal $h_k,p_k$. To see that this can be done we argue as follows (cf. \cite{BielCMP}). By the argument in the proof of Proposition \ref{bundle}, we can first solve the real equation on each $(\infty_{k-1},\mu_k)\cup(\mu_k,\infty_k)$ by bounded gauge transformation $g_k(t)$ satisfying the matching condition at $\mu_k$. Now, from Corollary \ref{C*} and the definition of $\tilde{M}$ as the hyperk\"ahler (and so complex-symplectic) quotient of the $\tilde{F}_{(n,m)}(c,c^\prime)$, we conclude that there is a global action of $({\Bbb C}^\ast)^m$ on $\tilde{M}$. Using this action allows us to replace the $g_k$ by a gauge transformation $g(t)$ which is exponentially close to $\exp\{h_kt-p_k\}$ near each $\infty_k$ and which also solves the real equation.
\par
We still have to show that $\phi$ respects complex-symplectic forms. However, $\phi$ was constructed using only a) complex gauge transformations, and b) restriction or extension of constant solutions. Both of these operations respect the complex-symplectic forms involved.\newline
{\bf Part 2: Estimates on solutions.}
We first obtain estimates on solutions to Nahm's equations. Recall that the biholomorphism $\phi$ was defined as the composition $\phi=\phi_2\phi_1$ with $\phi_1:\tilde{M}\rightarrow M_I$ and $\phi_2:M_I\rightarrow M$. Let $(\alpha,\beta)$ be a  solution to Nahm equations on a half-line $[x,+\infty)$, with $\min\{|\beta_{ii}(+\infty)-\beta_{jj}(+\infty)|; i\neq j\}\geq \delta>0$.
For any $\epsilon >0$, we can assume that $\alpha$ and $\beta$ are lower-triangular on $[x+\epsilon/2,+\infty)$ (this is done as in \cite{BielAGAG}: one can conjugate $\beta$ to be lower-triangular on $[x+\epsilon/2,+\infty)$ by a unitary gauge transformation; \eqref{complex} implies then that $\alpha$ is also lower-triangular). Then the apriori estimates from section 1 in \cite{BielAGAG} show that
\begin{equation}|\alpha_{ij}(t)|+|\beta_{ij}(t)|\leq Me^{-\lambda R_{ij}t}\label{betaij}\end{equation} 
for $i>j$ and $t\geq x+\epsilon$. Here $R_{ij}=|\real \alpha_{ii}(+\infty)- \real \alpha_{jj}(+\infty)|+|\beta_{ii}(+\infty)-\beta_{jj}(+\infty)|$ and $M,\lambda>0$ are constants depending only on $\delta,\epsilon$ (and the Lie-algebra to which $\alpha,\beta$ belong). For the diagonal part of $\alpha$ one has the following estimate (\cite{BielAGAG}, end of section 1):
$$|\real\alpha_{ii}(t)-\real\alpha_{ii}(+\infty)|\leq K$$
for all $i$ and $t\geq x+\epsilon$, $K=K(\delta,\epsilon)$. Then the real Nahm equation \eqref{real} gives 
$$\frac{d}{dt}(\real\alpha_{ii})\leq M\sum_{i>j} R_{ij} e^{-\lambda R_{ij}t},$$
from which we conclude, that for all $i$ and $t>x+\epsilon$,
\begin{equation}|\real\alpha_{ii}(t)-\real\alpha_{ii}(+\infty)|\leq M e^{-\lambda Rt},\quad R=\min R_{ij}.\label{alphaii}\end{equation}
Notice also that we can use the gauge freedom to make $\imag \alpha_{ii}$ constant on $[x+\epsilon,+\infty)$.
\par
 Now, $\phi_1$ was defined by a complex gauge transformation $p(t)$, with $p(\infty_k)=1$ and $p(t)=1$ on each $[\mu_k,\mu_k+c/2]\cup[\mu_{k+1}-c/2,\mu_{k+1}]$, making $\alpha$ and $\beta$ constant and diagonal on $[[\mu_k+c,\mu_{k+1}-c]]$. Thus we conclude that $(\tilde{\alpha}, \tilde{\beta})=\phi_1(\alpha,\beta)$ satisfies
\begin{equation}|\tilde{\alpha}_(t)-\alpha(t)|=\begin{cases} 0 & \text{if $t\in[\mu_k,\mu_k+c/2]\cup[\mu_{k+1}-c/2,\mu_{k+1}]$}\\ O\bigl(\exp\{-\lambda R_k t\}\bigr) & \text{if $t\in[[\mu_k+c,\mu_{k+1}-c]]$},\end{cases}\label{est1}\end{equation}
and similarly for $\beta$ and for the derivative of $\alpha$.
We now consider $\phi_2$. After cutting off the solutions, we obtain a solution $(\hat{\alpha},\hat{\beta})$ on $[\mu_1,\mu_N]$ to the complex Nahm equation which satisfies   $$F(\hat{\alpha},\hat{\beta}):=\frac{d\,}{dt}\left(\hat{\alpha}+ \hat{\alpha}^\ast\right)+[\hat{\alpha},\hat{\alpha}^\ast]+ [\hat{\beta},\hat{\beta}^\ast]=O(e^{-\lambda R}).$$
We know from the work of Hurtubise that there is a unique element of ${\cal G}^{\Bbb C}/{\cal G}$ such that any element $g(t)$ in this orbit takes $(\hat{\alpha},\hat{\beta})$ to an element of $M$. We have 
\begin{lemma} The gauge transformation $g$ satisfies $|g^\ast g-1|=O(e^{-\lambda R})$ uniformly on $[\mu_1,\mu_N]$.\label{gissmall}\end{lemma}
\begin{pf}
Using Lemma 2.10 in \cite{Don} and a simple comparison theorem (\cite{BielCrelle}, Lemma 2.8), one shows that the real equation can be solved on each $[\mu_k,\mu_{k+1}]$ by a complex gauge transformation $g_k(t)$ with $g_k(\mu_k)=g_k(\mu_{k+1})=1$ and $g_k^\ast g_k$ uniformly bounded by $O(e^{-\lambda R})$. Furthermore, near $\mu_k$, $|g_k^\ast g_k(t)-1|\leq (t-\mu_k)ce^{-\lambda R}$ and similarly near $\mu_{k+1}$. Therefore the derivative of $g_k^\ast g_k$  at $\mu_k,\mu_{k+1}$ is bounded by $ce^{-\lambda R}$. This shows that, while the resulting $\check{\alpha}$ does not satisfy the matching conditions at the $\mu_k$, the jumps are of order $O(e^{-\lambda R})$. Hurtubise shows in \cite{Hurt} that one can now match the solutions by a unique (complex) gauge transformation $g^\prime$. Since both $(\check{\alpha},\check{\beta})$ and $g^\prime (\check{\alpha},\check{\beta})$ satisfy the real equation, Lemma 2.10 in \cite{Don} implies that $g^{\prime\ast}  g^\prime$ is bounded by its values at the points $\mu_k$. Let $\phi$ (resp. $-\psi$) be the logarithm of maximum (resp. minimum) of eigenvalues of $g^{\prime\ast}  g^\prime$. The proofs of Propositions 2.20 and 2.21 in \cite{Hurt} show that the jumps $\Delta\dot{\phi}, \Delta\dot{\psi}$ of derivatives $\dot{\phi}$ and $\dot{\psi}$ are of order  $e^{-\lambda R}$ at each $\mu_k$. We then conclude, by going through the proof of Lemma 2.19 in \cite{Hurt}, that at each $\mu_k$ we have $\phi(\mu_k)\leq c \Delta\dot{\phi}+O(e^{-\lambda R})$ and $\psi(\mu_k)\leq c \Delta\dot{\psi}+O(e^{-\lambda R})$ for some $c<0$ (depending only on the $\mu_j$). This shows that $\phi(\mu_k)$ and $\psi(\mu_k)$ are both of order $e^{-\lambda R}$ which finishes the proof.\end{pf}

{\bf Part 3: Estimates for the tangent vectors.}
Recall that a tangent vector to a moduli space of solutions to Nahm's equations is a quadruple  $t_0,\dots t_3$ satisfying equations \eqref{tangent}. We shall write $a=t_0+it_1$ and $b=t_2+it_3$. Then the equations \eqref{tangent} can be written as
\begin{equation} \dot{a}=[\alpha^\ast,a]+[\beta^\ast,b],\label{treal}\end{equation}
\begin{equation} \dot{b}=[\beta,a]+[b,\alpha].\label{tcomplex}\end{equation}
If the moduli space consits of solutions defined on several adjoining intervals, then $a$ and $b$ also satisfy appropriate matching conditions at the endpoints.
\par
We shall need  apriori estimates for solutions of the above equations in $\tilde{F}_{n,m}(c,c^\prime)$. Let us write $\bigl((\alpha^-,\beta^-),(\alpha^+,\beta^+)\bigr)$ for a representative of $\tilde{F}_{n,m}(0,0)$ (and so of any $\tilde{F}_{n,m}(c,c^\prime)$) and then 
$x_i^-,z_i^-$ (resp. $x_i^+,z_i^+$) for the values of $\real\alpha^-,\beta^-$ (resp. $\real\alpha^+,\beta^+$)  at $-\infty$ (resp. at $+\infty)$). Let us also write $R^{\pm}=\min \{|x_i^{\pm}-x_j^{\pm}|+|z_i^{\pm}-z_j^{\pm}|;i\neq j\}$, $Z^\pm= \min \{|z_i^{\pm}-z_j^{\pm}|;i\neq j\}$ and $S=\min \{\{|x_i^{-}-x_j^{+}|+|z_i^{-}-z_j^{+}|\}$. We have
\begin{proposition} For any positive $\delta,\epsilon,\nu>0$ there exist constants $M,C,\lambda>0$ depending only on $m,n,\epsilon,\delta,\nu$ with the following property:
\par
Let $\bigl((\alpha^-,\beta^-),(\alpha^+,\beta^+)\bigr)$ be a representative of $\tilde{F}_{n,m}(0,0)$ with $Z^{\pm}\geq \delta>0$, $S\geq \nu >0$ and $R^+,R^->C$. If $((a^-,b^-),(a^+,b^+))$ is a tangent vector to $\tilde{F}_{n,m}(0,0)$ at $\bigl((\alpha^-,\beta^-),(\alpha^+,\beta^+)\bigr)$ and 
\begin{equation}A^2=|a^-(-\infty)|^2+|b^-(-\infty)|^2+|a^+(+\infty)|^2+ |b^+(+\infty)|^2,\label{A2}\end{equation}
then for all $t\geq \epsilon$
\begin{equation} |a^-(-t)-a^-(-\infty)|+|b^-(-t)-b^-(-\infty)| \leq Me^{-\lambda R^-t}A,\label{a-}\end{equation}
\begin{equation} |a^+(t)-a^+(+\infty)|+|b^+(t)-b^+(+\infty)| \leq Me^{-\lambda R^+t}A.\label{a+}\end{equation}\label{estab}\end{proposition}
\begin{pf} It is enough to prove the estimates for $A=1$. We can assume, as in part 2 of this proof, that $\alpha^\pm(t) , \beta^\pm(t)$ are lower-triangular for $|t|\geq \epsilon/2$. For the time being we consider only $\alpha^+ , \beta^+$ and we omit the superscript $+$.
We choose $C$ so that the right-hand side of \eqref{betaij} is small compared to $R_{ij}^{-1}$ and the right-hand side of \eqref{alphaii} is small compared to $R^{-1}$ at $t=\epsilon$. Then, if we write $y$ for the diagonal components  and $x$ for the off-diagonal components of  $a$ and $b$, we obtain from equations \eqref{treal} and \eqref{tcomplex}
\begin{equation} \dot{y}=A(t)x,\quad |A(t)|\leq Me^{-\lambda Rt}.\label{yyy}\end{equation}
On the other hand, if we differentiate the equations \eqref{treal} and \eqref{tcomplex}, we can write
\begin{equation} \ddot{x}=D(t)x+B(t)y,\quad  |B(t)|\leq Me^{-\lambda Rt},\enskip\exists_{s>0}\forall_z \real\bigl(D(t)z,z)\geq s^2R^2 |z|^2.\label{xxx}\end{equation}
Let $t_0\in[\epsilon,+\infty]$ be the first point for which $|x(t_0)|^2+|y(t_0)|^2\leq |a(+\infty)|^2+ |b(+\infty)|^2\leq 1$. 
Let  $X=\sup\{x(t);t\in [t_0,+\infty]\}$, $Y=\sup\{y(t);t\in [t_0,+\infty]\}$. Both $X$ and $Y$ are finite. Equation \eqref{yyy} implies that 
\begin{equation}|y(t)-y(t_0)|\leq \frac{MX}{R}.\label{yt}\end{equation}
 Similarly, using \eqref{xxx} and a comparison theorem (the same argument as on p. 133 in \cite{BielAGAG}), one concludes that
\begin{equation}|x(t)|\leq |x(t_0)|MYe^{-\lambda R(t-t_0)}\label{xt}\end{equation}
From this, changing $C$ if necessary (i.e. taking larger $R$), we conclude that there this a constant $P$ such that $X+Y\leq P$. Using again \eqref{yyy},\eqref{xt} we obtain that the estimate \eqref{a+} holds for $t\geq t_0$. This implies that  
\begin{equation}\int_{t_0}^{+\infty} \bigl(|a^+(t)|^2+|b^+(t)|^2- |a^+(+\infty)|^2-|b^+(+\infty)|^2\bigr)dt\leq \rho,\label{fromt0}\end{equation}
where $\rho=\rho(n,\delta,\epsilon)$ and can be made arbitrarily small by changing $C$ (recall that $A=1$). We also have that for $t\in[\epsilon,t_0]$ the expression under the integral sign in \eqref{fromt0} is nonnegative. 
We can do exactly the same for $\alpha^-,\beta^-,a^-,b^-$. Let $s_0$ denote the negative number with the same properties as $t_0$. We now compute the length $L$ of the vector $\bigl((a^-,b^-),(a^+,b^+)\bigr)$ in the metric of $\tilde{F}_{(n,m)}(\epsilon,\epsilon)$. We can write (cf. \eqref{smetric}) (the fact that below we have an inequality, rather than equality, stems from the fact that, for $n=m$, there are additional (positive) terms):  
\begin{multline}
L^2\geq\int_{-\epsilon}^0\bigl(|a^-(t)|^2+|b^-(t)|^2\bigr)dt +\int_0^\epsilon \bigl(|a^+(t)|^2+|b^+(t)|^2\bigr)dt\\
+\int^{-\epsilon}_{s_0}\bigl(|a^-(t)|^2+|b^-(t)|^2- |a^-(-\infty)|^2-|b^-(-\infty)|^2\bigr)dt \\+\int^{t_0}_\epsilon \bigl(|a^+(t)|^2+|b^+(t)|^2- |a^+(+\infty)|^2-|b^+(+\infty)|^2\bigr)dt\\ +\int_{-\infty}^{s_0}\bigl(|a^-(t)|^2+|b^-(t)|^2- |a^-(-\infty)|^2-|b^-(-\infty)|^2\bigr)dt\\ +\int_{t_0}^{+\infty} \bigl(|a^+(t)|^2+|b^+(t)|^2- |a^+(+\infty)|^2-|b^+(+\infty)|^2\bigr)dt. \label{longformula}\end{multline}
Each of the first four lines is positive, while the last two have their absolute value bounded by $2\rho$ with $\rho$ as small as we wish. Let us write $T$ for the sum of the third and fourth line. It follows that $T\leq L^2+2\rho$. The  explicit formula for the metric on $ \tilde{F}_{n,m}(c,c^\prime)$ found in sections \ref{five} and \ref{six} implies that $L^2\leq P$, where $P>0$ depends only on $m,n,\epsilon,\nu,\delta$ (notice that this bound is independent of the actual value of $\epsilon_{ij}$ in \eqref{sij}). Thus $T\leq P^\prime$. Now, if both $t_0$ and $-s_0$ are smaller than $2\epsilon$, then we are done (by replacing the original $\epsilon$ with $2\epsilon$). Suppose that $t_0\geq 2\epsilon$. Since the integrand in the second and third line is nonnegative, we conclude from  $T\leq P^\prime$ that there is point $t_1\in [\epsilon,2\epsilon]$ with $|a^+(t_1)|^2+|b^+(t_1)|^2\leq P^\prime/\epsilon$. We can now repeat the arguments after \eqref{xxx} and conclude that the estimate \eqref{a+} holds for $t\geq 2\epsilon$. We can deal similarly with the case $s_0\leq -2\epsilon$.
\end{pf}

We shall also need the following strengthening of the last result:
\begin{lemma} With the same assumptions and notation as in Proposition \ref{estab}, we can replace the estimates \eqref{a-} and \eqref{a+} with:
$$ |a_{ij}^-(-t)|+|b_{ij}^-(-t)| \leq Me^{-\lambda R_{ij}^-t}A,\qquad
 |a_{ij}^+(t)|+|b_{ij}^+(t)| \leq Me^{-\lambda R_{ij}^+t}A,$$
for all $i\neq j$ and $t\geq\epsilon$. Here $R_{ij}^{\pm}=|x_i^\pm -x_j^\pm|+ |z_i^\pm -z_j^\pm|$.\label{expotan}\end{lemma}
\begin{pf} We differentiate the equations \eqref{treal} and \eqref{tcomplex} and proceed, as in Proposition 3.12 of \cite{BielCrelle} using  \eqref{a-} and \eqref{a+}.\end{pf}

{\bf Part 4: Proof of Theorem \ref{finalasymptoticmetric}.} From its definition in section \ref{one}, $F_{\tilde{m}_1,\dots,\tilde{m}_{N-1}}(\mu_1,\dots,\mu_N)$ is the hyperk\"ahler quotient, by a product of tori, of the product 
$$\tilde{F}_{m_1}(c_1)\times \tilde{F}_{m_2,m_1}(c_2,c_2^\prime ) \times\dots\times \tilde{F}_{m_{N-1},m_{N-2}}(c_{N-1},c_{N-1}^\prime)\times \tilde{F}_{m_N}(c_N^\prime)$$
where $c_i+c_{i+1}^\prime=\mu_{i+1}-\mu_{i}$, $i=1,\dots,N-1$. The matrices $\Phi$ for each factor are of the form \eqref{GMtype} with the $s_{ij}$ given by \eqref{sij} (for the first and last factor, the metric is the Gibbons-Manton metric given by \eqref{GM}). On the hyperk\"ahler quotient these matrices are simply added together (after viewing each of them as a submatrix of an $m\times m$ matrix, $m=m_1+\dots+m_{N-1}$). Thus the result is proved as soon as we show that all $\epsilon_{ij}$ of \eqref{sij} are zero. Let us show this.
\begin{lemma} In the formula \eqref{sij}, all $\epsilon_{ij}$ are equal to zero.\label{epsilonij}\end{lemma}
\begin{pf}
Suppose that this is not true. Let us write the norm of vector $v$ tangent to $F_{(n,m)}(1,1)$ as in the formula \eqref{longformula} with $\epsilon=t_0=-s_0=1$. From the estimates \eqref{a-} and \eqref{a+}, it follows that $\|v\|^2\geq -MA^2/R$ for a constant $M$ depending only $m,n,\delta$ and $\nu$. Here $R=\min\{R^-,R^+\}$.  Thus, for a sufficiently large $R$, $\|v\|^2\geq -\rho A^2$, $A$ defined by \eqref{A2}, with $\rho$ as small as we wish. However, if any $\epsilon_{ij}=1$, then we can find a point ${\bf x}$ in  $F_{(n,m)}(1,1)$ with $S=S({\bf x})\geq \nu$ and $R$ arbitrarily large such that there is a tangent vector $v$ at ${\bf x}$ with $\|v\|^2\leq -c A^2/\nu$ for some $c=c(n,m)$. This contradicts $\|v\|^2\geq -\rho A^2$ and so the lemma is proved.\end{pf}

{\bf Part 5: Proof of Theorem \ref{estimates}.}

In appendix B we prove a general theorem we allows us to reduce the estimates to one-sided estimates on the metric tensors. This is so because the asymptotic metric is quasi-isometric to the flat metric in coordinates \eqref{cancoor}. This last fact follows from the explicit formula for the metric and the twistor space (cf. Remark 2 after Theorem \ref{estimates}). 
Thus we only have to show that
\begin{equation} \phi^\ast g\leq \bigl(1+Me^{-\lambda R}\bigr)\tilde{g}\label{estXX}\end{equation}
in the region $U(\gamma,\delta,C)$ for some $C,M,\lambda>0$ depending only on $\gamma,\delta$. Once we have this, we apply Theorem \ref{B1} (and Remark \ref{B2}) to the region where $R\geq R_0$ and obtain that the estimate \eqref{DkSij} with $R=R_0$, $R_0$ arbitrary, holds in the region where $R\geq R_0+1$, in particular for all points with $R=R_0+1$. Since $R_0$ is arbitrary this will prove the theorem.
\par
Therefore we are going to show \eqref{estXX}. We start with a vector $(a,b)$ tangent to $U(\gamma,\delta,C)$, where $C=C(\gamma,\delta)$ is determined by the validity of estimates below. Since $\gamma>0$, the metric is positive-definite and, furthermore, quasi-isometric to the flat metric in coordinates \eqref{cancoor} or the coordinates given by positions and phases of particles. Let us assume that the norm of $(a,b)$ is $1$ in this metric. Then $\sum_k \bigl(|a(\infty_k)|^2+|b(\infty_k)|^2\bigr)\leq B$, where $B$ depends only on $\gamma$. We also have estimates of the form \eqref{a-} and \eqref{a+}:
\begin{equation} |a(t)-a(\infty_k)|+|b(t)-b(\infty_k)|\leq Me^{-\lambda R_k t}B\quad \text{if $t\in [[\mu_k+\epsilon,\mu_{k+1}-\epsilon]]$}, \label{a-infty}\end{equation}
as well as the stronger estimates of Lemma \ref{expotan}.
Also, by writing the metric as in \eqref{longformula} with $\epsilon=t_0=c$ and using \eqref{a-infty}, we get
\begin{equation} \sum_{k=1}^{N}\int_{\mu_k-c}^{\mu_k+c} \bigr(|a(t)|^2+|b(t)|^2\bigr)dt \leq MB.\label{L2a}\end{equation}
The left-hand side includes the sum of Euclidean norms of pairs of vectors $u_k,v_k$ which give us the matching conditions for $(a,b)$ at $\mu_k$ in the case when $m_{k-1}=m_k$.
\par 
Recall that the map $\phi$ was a composition of a $\phi_1$ and $\phi_2$. The map $\phi_1$ was given by a complex gauge transformation $p(t)$ which from part 2 of the proof can be uniformly estimated by $O(e^{-\lambda Rt})$. Therefore, after we conjugate $a,b$ by $p$, they still satisfy \eqref{a-infty}. Moreover we have $\bigl|\|(pap^{-1},pbp^{-1})\|-1\bigr|=O(e^{-\lambda R})$. In order to obtain the vector $d\phi_1\bigr((a,b)\bigr)$, one has to make $pap^{-1}$ and $pbp^{-1}$ constant and diagonal on each $[[\mu_k+c,\mu_{k+1}]]$ ($c$ is defined in part 1) by an infinitesimal complex gauge transformation $\rho_1$ (with $\rho_1(\infty_k)=0$ etc.). From the estimates \eqref{a-infty} and of Lemma \ref{expotan} on $pap^{-1}$ and $pbp^{-1}$, this changes the norm of $p(a,b)p^{-1}$ by something of order $e^{-\lambda R}$. Furthermore the $L^2$-estimate \eqref{L2a} holds for $d\phi_1(a,b)$. At the next stage, we restrict $d\phi_1(a,b)$ to $[\mu_1,\dots,\mu_N]$. Since $d\phi_1(a,b)$ is constant and diagonal on the union of $[[\mu_k+c,\mu_{k+1-c}]]$, its norm in the metric $\tilde{g}$ is the same as the norm of the restriction $(\hat{a},\hat{b})$ in the metric $g$. Now we conjugate  $(\hat{a},\hat{b})$ by the complex gauge transformation $g(t)$ of Lemma \ref{gissmall}. Using the estimate of that lemma and the estimate \eqref{L2a} for $(\hat{a},\hat{b})$ we conclude that
\begin{equation} \bigl|\|(g\hat{a}g^{-1},g\hat{b}g^{-1})\|-1\bigr|\leq Me^{-\lambda R},\label{almostthere}\end{equation}
for some $M,\lambda>0$ depending only on $\gamma$ and $\delta$. The vector $(g\hat{a}g^{-1},g\hat{b}g^{-1})$ solves the equation \eqref{tcomplex} but not \eqref{treal}. This is the final step: we obtain the vector $d\phi(a,b)$ by acting on $(g\hat{a}g^{-1},g\hat{b}g^{-1})$ with a complex infinitesimal gauge transformation, so that the resulting vector solves \eqref{treal}. However, the equation \eqref{treal} is the condition of orthogonality to complex infinitesimal gauge transformations and, hence, the norm of the vector $d\phi(a,b)$ is not greater than the norm of $(g\hat{a}g^{-1},g\hat{b}g^{-1})$. This and \eqref{almostthere} proves \eqref{estXX}, and so, by the discussion above, also Theorem \ref{estimates}\hfill $\Box$\bigskip

As remarked after the statement of the above theorem, there is a likely generalization of this result. Suppose that it is only particles of a given type, say $k_0$, that separate (recall that the type of particle $i$ is the smallest $k$ for which $i\leq m_k$). Then the metric on  $F_\sigma(\mu)=F_{m_1,\ldots,m_{N-1}}(\mu_1,\ldots,\mu_N)$) should get close to the metric on $F_{\tilde{\sigma}}(\mu)$, where $\tilde{\sigma}(k)=m_k$ if $k\neq i_0$ and $\tilde{\sigma}(k_0)=\tilde{m}_{k_0}$. Similarly, if the particles of types $k_1,\dots,k_s$ separate, the metric should be close to the metric on $F_{\tilde{\sigma}}(\mu)$, where $\tilde{\sigma}(k)=m_k$ if $k\neq k_1,\ldots,k_s$ and $\tilde{\sigma}(k_j)=\tilde{m}_{k_j}$, for $j=1,\ldots,s$. All of these moduli spaces have dimension $4(m_1+\ldots+m_{N-1})$. In general the metric on $F_{\tilde{\sigma}}(\mu)$ will be simpler than the one on $F_\sigma(\mu)$ (it has a tri-Hamiltonian action of a $(\sum_{i=1}^s m_{k_i})$-dimensional torus), but it is only in the case when $\{k_1,\ldots,k_s\}=\{1,\ldots,N-1\}$ that the metric is algebraic.

Finally we shall discuss the topology of the asymptotic moduli space. First of all, from Proposition \ref{bundle}, the orbit space of $F_{\tilde{m}_1,\dots,\tilde{m}_{N-1}}(\mu_1,\dots,\mu_N)$ is $\prod \tilde{C}_{m_k}\bigl({\Bbb R}^3)$, and so particles of different types can take the same position. Now recall from section \ref{zero} that the type $t(i)$ of the particle $i$ is defined as $\min\{k;i\leq \sum_{s\leq k}m_s\}$. It follows easily from Proposition \ref{bundle} that the set of principal orbits of $T^m$, $m=m_1+\dots+m_{N-1}$, on $F_{\tilde{m}_1,\dots,\tilde{m}_{N-1}}(\mu_1,\dots,\mu_N)$ is a bundle $P$ over 
$$C=\bigl\{({\bf x}_1,\dots,{\bf x}_m)\in{\Bbb R}^3\otimes {\Bbb R}^m; |t(i)-t(j)|\leq 1\implies {\bf x_i}\neq {\bf x_j}\bigr\}.$$
The basis of the second integer homology of $C$ is given by the spheres $S_{ij}$ defined by \eqref{basis}, where now $i,j$ run over the set $\{(i,j);i<j\enskip \text{and}\enskip |t(i)-t(j)|\leq 1\}$. As in Lemma 7.1 in \cite{BielCMP}, we obtain that the bundle $P$ is determined by the element $(h_1,\ldots,h_{m})$ of $H^2\bigl(C,{\Bbb Z}^{m}\bigr)$ such that
$$h_k(S_{ij})=\begin{cases}s_{ij} & \text{if $k=i$}\\ -s_{ij} & \text{if $k=j$} \\ 0 & \text{otherwise},\end{cases}$$
where the $s_{ij}$ are given by \eqref{s-asym}.

\appendix
\section{Complex-symplectic vs. hyperk\"ahler quotients}

In order to finish the proof of Proposition \ref{bundle} we have to show that certain hyperk\"ahler and complex-symplectic quotients coincide. This question reduces, on the zero-set of the complex moment map, to identifying the K\"ahler quotient with the ordinary geometric quotient, i.e. to showing that all orbits of the complexified group are stable. The following useful criteria are given in \cite{HH}, Lemma 3.3, and \cite{Gu}, Sections A.1.3 and A.2.3.
\begin{proposition} Let $H$ be a connected closed subgroup of a compact semisimple Lie group $G$ and suppose that  $M=G^{\Bbb C}/H^{\Bbb C}$ is equipped with a $G$-invariant K\"ahler form $\omega$ defined by a global $G$-invariant K\"ahler potential $K$ for $\omega$. Then the single $G^{\Bbb C}$-orbit $M$ is stable if and only if $K$ is proper.\label{stable1}\end{proposition}
\begin{proposition} Suppose that the complex torus $M=({\Bbb C}^\ast)^n$ is equipped with  a $T^n$-invariant K\"ahler form $\omega$. Then:
\begin{itemize}
\item[(i)] The $T^n$-action is Hamiltonian if and only if there exists a global $T^n$-invariant K\"ahler potential $K$ for $\omega$;
\item[(ii)] If $K$ has a quadratic growth at infinity, as a function  on ${\Bbb R}^n$ where $({\Bbb C}^\ast)^n=T^ne^{{\Bbb R}^n}$, then $M$ is stable. \end{itemize}
\label{stable2}\end{proposition}

In order to use these criteria we shall view the spaces involved in the proof of Proposition \ref{bundle} as hyperk\"ahler quotients of simpler manifolds.
\par
Recall that, for a regular triple $(\tau_1,\tau_2,\tau_3)$ of $n\times n$ diagonal matrices, $M(\tau_1,\tau_2,\tau_3)$ denotes Kronheimer's hyperk\"ahler structure on $Gl(n,{\Bbb C})/T^{\Bbb C}$ ($T$ is the diagonal torus in $U(n)$) with the cohomology class of $\omega_i$ equal to $\tau_i$, $i=1,2,3$ (after identifying $H^2\bigl(Gl(n,{\Bbb C})/T^{\Bbb C}, {\Bbb R}\bigr)$ with $\text{Lie}(T)$). We have
\begin{proposition}  $M(\tau_1,\tau_2,\tau_3)$ is isomorphic to a hyperk\"ahler quotient of a flat quaternionic vector space by a product $G$ of unitary groups. With respect to generic complex structure the action of $G^{\Bbb C}$ on the zero-set of the complex moment map is free and its orbits are closed.\label{orbits}\end{proposition}
\begin{pf}  Kobak and Swann \cite{KS} show how to construct nilpotent orbits as hyperk\"ahler quotients by a product $G$ of unitary groups. Changing the level set of the moment map, from zero to appropriate values, gives a hyperk\"ahler manifold $N$ which, with respect to a generic complex structure, is isomorphic, as a complex-symplectic manifold, to  $M(\tau_1,\tau_2,\tau_3)$ (if the levels of the complex moment map for the abelian
factors of $G$ are $t_1,\dots,t_{n-1}$, then the resulting orbit has eigenvalues $0,t_1,t_1+t_2,\dots,(t_1+\dots+t_{n-1})$). As a Riemannian manifold, $N$ is complete and the uniqueness result of \cite{BielCam} shows that $N\simeq M(\tau_1,\tau_2,\tau_3)$ as hyperk\"ahler manifolds. 
\par
The flat space which we start with can be viewed as a space of matrices and   the second statement follows by putting the products of matrices defining the zero set of the complex moment map into the Jordan normal form.\end{pf} 

We can finally finish the proof of Proposition \ref{bundle}. We considered there a hyperk\"ahler quotient $X$ of the product of two $M(\tau_1,\tau_2,\tau_3)$'s and of either ${\Bbb H}^n$ or of $F_n(m;1)$. By the above proposition, $X$ can be viewed as a hyperk\"ahler quotient of either an ${\Bbb H}^p$ (when $n=m$), for some $p$, or of the product of ${\Bbb H}^p$ and of $F_n(m;1)$ (when $n>m$). Furthermore, the complexification $H^{\Bbb C}$ of the group $H$ by which we quotient acts freely and with closed orbits on the zero-set of the (generic) complex moment map. In addition the usual K\"ahler potential $K_1$, given by the square of the distance from the origin, on ${\Bbb C}^{2p}\simeq {\Bbb H}^p$ is  proper and $Sp(p)$-invariant. Moreover it has quadratic (in fact exponential) growth on any closed orbit of a subtorus of $Sp(p,{\Bbb C})$.

Since the $H^{\Bbb C}$-orbits are closed on the zero set of the complex moment map, Propositions \ref{stable1} and \ref{stable2} show that all these $H^{\Bbb C}$-orbits are stable for $n=m$.
\par
The space $F_n(m;1)$ also has an $\bigl(U(n)\times U(m)\bigr)$-invariant K\"ahler potential for any complex structure. This is a general phenomenon for hyperk\"ahler manifolds with an $SU(2)$-action rotating the complex structures, see \cite{HKLR}. For the complex structure $I$ this K\"ahler potential is given by:
$$K=\int_0^1\bigl(\|T_1(t)\|^2+\|T_2(t)\|^2\bigr)dt.$$
Since the factor $\bigl(U(n)\times U(m)\bigr)$ of $H$ is acts diagonally on ${\Bbb H}^p\times F_n(m;1)$, the K\"ahler potential $K_2$ on each orbit of $Gl(n,{\Bbb C})\times Gl(m,{\Bbb C})$ is the sum of the restrictions of $K_1$ and $K$. Since $K_1$ is proper and has quadratic growth on each closed toral orbit and since $K$ is positive, it follows that $K_2$ is proper (on each orbit) and has quadratic growth on each closed toral orbit. Thus the complex-symplectic quotient coincides with the hyperk\"ahler quotient in this case as well.

\section{A comparison theorem for Ricci-flat K\"ahler metrics}

Our goal is the following theorem, which, under certain assumptions, reduces the comparison of Ricci-flat K\"ahler metrics and their derivatives to one-sided estimates on the metric tensors.
\begin{theorem}  Let $V$ be an open subset of a Ricci-flat K\"ahler manifold $(M,g)$ and suppose that there exists a biholomorphic map $\phi$ from a domain $U$ in ${\Bbb C}^m$ onto $V$ 
such that $\phi^\ast g$ is bounded uniformly in the Euclidean metric on $U$ by 
a constant $C$ and that $\phi^\ast \omega^m=e^f\omega_0^m$ with $f$ bounded uniformly by a constant $K$  
(here $\omega,\omega_0$ are the K\"ahler forms on $X$ and on ${\Bbb C}^n$). Then, for any $r,\delta>0$, there exist constants $A_k=A_k(m,C,K,r,\delta)$, $k=0,1,2,\dots$, 
with the following property:
\par
Let $(M^\prime,g^\prime)$ be another K\"ahler manifold and $\psi:M\rightarrow M^\prime$ a volume-form preserving biholomorphism such that $\psi^\ast g^\prime\leq (1+\epsilon)g$, uniformly on $V$, where $\epsilon\leq \delta$. Let us write  $\phi^\ast (\psi^\ast g^\prime -g)= \real\sum S_{ij}dz_i\otimes d\bar{z}_j$. Then, for any $i,j\leq m$ and any $k\geq 0$,
\begin{equation} |D^{k}S_{ij}|\leq A_k\epsilon\label{MAest2}\end{equation}
uniformly on the set $\{z\in U; \text{dist}\,(z,\partial U)\geq r\}$.\label{B1}\end{theorem}
 
\begin{remark} In hyperk\"ahler geometry the conditions on the volume-form are very natural. Namely, if $M$ is hyperk\"ahler with the complex-symplectic form $\Omega=g(J\,,\,)+ ig(IJ\,,\,)$ and $\phi^\ast \Omega$ is the standard complex-symplectic form on ${\Bbb C}^{2n}$, then $\phi^\ast \omega^{2n}=\omega^{2n}_0$, so that we can take $K=0$ in the theorem. Similarly, if $M^\prime$ is hyperk\"ahler and $\psi$ respects the complex-symplectic forms, then $\psi$ respects the volume forms.\label{B2}\end{remark}
We also remark that the assumption that $\phi^\ast g$ and $\ln(\omega^m/\omega^m_0)$ are uniformly bounded is equivalent to $\phi$ being a quasi-isometry, i.e. to existence of a constant $B$ such that $B^{-1}g_0\leq \phi^\ast g \leq Bg_0$, where $g_0$ is the Euclidean metric on ${\Bbb C}^m$. The fact that one-sided estimates on the metric plus estimates on the volume form give two-sided estimates on the metric is trivial but, given Remark \ref{B2}, worth stating separately:
\begin{proposition} Let $(M,g),(M^\prime,g^\prime)$ be two oriented Riemannian manifolds of dimension $n$ and let $\psi:M\rightarrow M^\prime$ be a volume-form preserving diffeomorphism such that $\phi^\ast g^\prime\leq Bg$. Then $B^{1-n}g\leq \phi^\ast g^\prime$.\label{B3}\end{proposition}
\begin{pf} Choose a point $m\in M$ and local coordinates $x_1,\dots,x_m$ so that $g_m=\sum dx_i\otimes dx_i$. Write $(\phi^\ast g^\prime)_m=\sum X_{ij}dx_i\otimes dx_j$. Then the assumptions imply that $\det X=1$ and $\zeta^TX\zeta\leq B$ for all $\zeta\in {\Bbb R}^n$. Since $X$ is symmetric and positive-definite we can diagonalize $X$ by an orthogonal matrix and conclude that $\zeta^TX\zeta\geq B^{1-n}\zeta^T\zeta$ for all $\zeta\in {\Bbb R}^n$.\end{pf} 
{\bf Proof of Theorem \ref{B1}.} We write 
\begin{equation}\phi^\ast g=\real\sum_{i,j=1}^m Z_{ij}dz_i\otimes d\bar{z}_j,\qquad \phi^\ast\psi^\ast g^\prime=\real\sum_{i,j=1}^m Z^\prime _{ij}dz_i\otimes d\bar{z}_j\label{pull}\end{equation}
for hermitian $Z,Z^\prime$.
Let us fix an $r$, so that we have to estimate the derivatives at points $z$ such that $B(z,r)\subset U$. We can assume, without loss of generality, that $U=B(0,r)$ and estimate the derivatives at the origin. Since $B(0,r)$ is strictly pseudo-convex, there are smooth real-valued functions $\Phi,\Phi^\prime$ such that $\Phi_{z_i\bar{z}_j}=Z_{ij}$ and  $\Phi^\prime _{z_i\bar{z}_j}=Z_{ij}^\prime$. In general, we shall write $L(u)$ for the complex Hessian (Levi form) $\bigl[u_{z_i\bar{z}_j}\bigr]$ of a function $u$. Both $\Phi$ and $\Phi^\prime$ satisfy the complex Monge-Amp\`ere equation
\begin{equation} \det L(u)=e^f\label{MA}.\end{equation}
Since $M$ is Ricci-flat, $f$ is a polyharmonic function. In particular all derivatives of $g=e^f$ have uniform bounds depending only on the bound $K$ for $f$. We need the following
\begin{proposition} Let $u$ be a pluri-subharmonic solution to \eqref{MA} in $B(0,r)$. Then, for any $k\geq 0$,
$$ \|L(u)\|^\ast_k\leq B_k,$$
where $B_k$ depends only on $m,r$, $\|L(u)\|_0$ and $\|f\|_{k+2}$. \label{B4}\end{proposition}
 Here $\|\:\|_k$ and $\|\:\|^\ast_k$ denote, respectively, the $C^k$-norm and the interior $C^k$-norm (see \cite{GT}, formula (4.17), for the definition of the latter).
\begin{pf} Let $B=\|L(u)\|_0$. Since $\det L(u)=e^f$, we have (as in Proposition \ref{B3}) positive constants $\lambda$ and $\Lambda$ depending only on $m,B$ and $\|f\|_0$ such that 
$$\lambda\zeta^\ast\zeta\leq \zeta^\ast L(u)\zeta\leq \Lambda\zeta^\ast\zeta$$
for all $\zeta\in {\Bbb C}^m$. Thus the equation \eqref{MA} is uniformly elliptic with respect to $u$. 
Using the estimates on a solution of $\bar{\partial}$-problem (see, for example, \cite{XX}), we see that there is a smooth real-valued
function $u^\prime$ with $L(u^\prime)=L(u)$ and $\|u^\prime\|_0\leq C\|L(u)\|_0$ for a constant $C=C(r)$. Therefore, we can assume from the beginning that $u$ is bounded by some $C_0$ depending only on $m,r$ and $\|L(u)\|_0$. We now claim that there are constants $C_1$ and $\alpha>0$, depending only on $m,r,\Lambda, \|f\|_2$ such that the following H\"older estimate holds
\begin{equation} \|L(u)\|_\alpha^\ast\leq C_1\end{equation}
The proof of this is essentially the same as the proof of (17.41) in \cite{GT} (applied to the equation $\log \det L(u)=f$), but using the Hermitian analogue of Lemma 17.13 in \cite{GT}. Once we have this, we obtain inductively estimates on all derivatives of $u$ by treating the equation \eqref{MA} and its successive linearizations in all directions as uniformly elliptic second-order linear PDE's (with the coefficients of second-order derivatives given by the adjoint matrix of $L(u)$). The desired estimates follow from standard Schauder interior estimates. \end{pf}

To finish the proof of Theorem \ref{B2}, we write the equation $\det L(\Phi)=\det L(\Phi^\prime)$ as 
\begin{equation}F\bigl(L(\Phi-\Phi^\prime)\bigr)=0\label{linear}\end{equation}
for a linear function $F$. The coefficients of $F$ depend only on $L(\Phi)$ and $L(\Phi^\prime)$. The function $F$ is uniformly elliptic in $B(0,3r/4)$ with ellipticity constants
depending only on $m,r,C,K,\delta$. Moreover, by the previous proposition, the derivatives of coefficients also have bounds depending only on these constants.
From Proposition \ref{B3}, $v=\Phi-\Phi^\prime$ satisfies $L(v)\leq P\epsilon$ for some constant $P$, and, as in the proof of Proposition \ref{B4}, we can use estimates (e.g. \cite{XX}) on the solution of  $\bar{\partial}$-problem to conclude that there is a $v^\prime$ with $L(v^\prime)=L(v)$ and  $\|v^\prime\|_0\leq P^\prime\epsilon$. We now obtain the estimates on all derivatives of $v^\prime$ from the Schauder interior estimates applied to the equation \eqref{linear} and its successive linearizations in all directions. \hfill $\Box$

\addtolength{\textheight}{1cm}

\end{document}